\documentclass[runningheads]{llncs}

\usepackage{graphicx}
\usepackage{amsmath}
\usepackage{amssymb}
\usepackage{url}
\usepackage{xcolor}
\usepackage{multirow}
\usepackage{hyperref}
\usepackage{subcaption}
\usepackage{acro}
\usepackage{cite}
\DeclareMathOperator*{\argmin}{arg\,min}
\newcommand\norm[1]{\left\lVert#1\right\rVert}
\begin{document}

\title{A Multiparametric Volumetric Quantitative Ultrasound Imaging Technique for Soft Tissue Characterization}
\titlerunning{Quantitative Ultrasound Imaging for Soft Tissue Characterization}
%
\author{Farah Deeba\inst{1} \and
Caitlin Schneider \inst{1}
\and
Mohammad Honarvar\inst{2}
\and
Julio Lobo\inst{2}
\and
Edward Tam\inst{3}
\and
Robert Rohling\inst{1,4}
\and 
Septimiu Salcudean\inst{1}
}
\authorrunning{F. Author et al.}
%
\institute{Department of Electrical and Computer Engineering,
The University of British Columbia, Vancouver, BC, Canada \\
\and
Sonic Incytes, Vancouver, Canada
\and
The Lair Centre, Vancouver, BC, Canada \and
Department of Mechanical Engineering,
The University of British Columbia, Vancouver, BC, Canada 
}
\maketitle              
\begin{abstract}
Quantitative ultrasound (QUS) offers a non-invasive and objective way to quantify tissue health. We recently presented a spatially adaptive regularization method for reconstruction of a single QUS parameter, limited to a two dimensional region. That proof-of-concept study showed that regularization using homogeneity prior improves the fundamental precision-resolution trade-off in QUS estimation. Based on the weighted regularization scheme, we now present a multiparametric 3D weighted QUS (3D QUS) imaging system, involving the  reconstruction of three QUS parameters: attenuation coefficient estimate (ACE), integrated backscatter coefficient (IBC) and effective scatterer diameter (ESD). With the phantom studies, we demonstrate that our proposed method accurately reconstructs QUS parameters, resulting in high reconstruction contrast and therefore improved diagnostic utility. Additionally, the proposed method offers the ability to analyze the spatial distribution of QUS parameters in 3D, which allows for superior tissue characterization. We apply a three-dimensional total variation regularization method for the volumetric QUS reconstruction. The 3D regularization involving $N$ planes results in a high QUS estimation precision, with an improvement of standard deviation over the theoretical $1/\sqrt{N}$ rate achievable by compounding $N$ independent realizations. In the \textit{in vivo} liver study, we demonstrate the advantage of adopting a multiparametric approach over the single parametric counterpart, where a simple quadratic discriminant classifier using feature combination of three QUS parameters was able to attain a perfect classification performance to distinguish between normal and fatty liver cases.

\end{abstract}

\section{Introduction}
\label{sec1}

The role of ultrasound, as a portable triage and monitoring tool, is becoming increasingly important, especially in the current context of a global pandemic. Conventional ultrasound images are formed from backscattered radio frequency echo signals created as a result of interaction with macro- and micro scale tissue structures. The B-mode image processing, including envelope detection and log compression, highlights the large-scale ($>$ultrasound wavelength) features such as the organ boundaries, while discarding the frequency-dependent smaller scale ($<$ultrasound wavelength) structural information. The speckle pattern and echogenecity of the B-mode image, though related to the complex organization of smaller scale tissue structure and their averaging scattering strength, render the US imaging qualitative, system-dependent and open to user interpretation. Three decades of research in the field of quantitative ultrasound (QUS), followed by the pioneering works by Insana et al. \cite{Gilmore2008} and Lizzi et al. \cite{ColemanD.Jackson1983}, established system and user independent techniques to quantify acoustic and microstructural properties by parametrizing the frequency dependent contents in the backscattered echo. Due to the reduced dependence on user interpretation as well as the efficacy for tissue typing and monitoring disease progression  substantiated by extensive experimental research \cite{Mamou, Oelze2016}, QUS imaging promises to further enhance the clinical utility of ultrasound. 

Spectral based parametric imaging forms a major branch of QUS, which includes measurement of attenuation coefficient estimate (ACE), backscatter coefficient and scatterer properties such as effective scatterer diameter (ESD) and effective acoustic concentration (EAC) of scatterers. ACE is a frequency dependent acoustic property that measures the ultrasound energy loss with propagation depth due to the combined effect of scattering and absorption. Backscatter coefficient, on the hand, measures the ultrasound intensity that is scattered in the backward direction (normalized ultrasound backscattered power per unit volume of scatterers). The integrated backscatter coefficient (IBC) is a measure of the backscatter strength, which is  estimated by integrating the backscatter coefficient in the effective frequency band of the transducer. Backscatter coefficient can further be used to estimate the ESD of the dominant sub-resolution ($<$wavelength) scatterer, with the effective scale being determined by the incident ultrasound frequency. The spectral based QUS parameters have been successfully used in numerous clinical applications, including cancer detection in prostrate and lymph nodes \cite{Feleppa11}, bone health assessment \cite{Guillaume2011}, cervical ripening detection \cite{Timothy12}, breast lesion characterization \cite{Nam2013}, and  placenta characterization \cite{Deeba2019}.

 Liver steatosis quantification using QUS has undergone a recent surge in an effort to address the alarming increase in non-alcoholic fatty liver disease prevalence rate (25\% globally) \cite{Loomba2013}. Previous works demonstrated that the cellular ballooning due to fat infiltration in the fatty liver affects the ultrasonic scattering process, resulting in an increase in ACE and Backscatter coefficient \cite{Pohlhammer1980, bamber1981}. Several clinical studies on QUS showed promising results to quantify the fat content in liver, with efficacy comparable to magnetic resonance imaging proton density fat fraction (MRI-PDFF) and biopsy, the current gold standard for steatosis grading \cite{ Andre2014, Lin2015, fujiwara2018, Deeba2019a}. 

Current challenges towards the clinical translation of QUS imaging include large estimation bias and variance due to tissue inhomogeneity and precision-resolution trade-off \cite{Liu2010, Oelze2004}. The resolution-precision trade-off is particularly critical for thin (e.g. dermis) and heterogeneous (e.g. placenta) tissue characterization. Available smaller windows result in noisy power spectral estimates due to limited spectral resolution and spatial variation noise inherent in ultrasound scattering. Deviation from the homogeneity assumption arising from the variation in underlying scatterer statistics \cite{pawlicki2013}, on the other hand, is likely to affect most biological tissue to different extent.  Rubert et al. demonstrated that the high variance (with standard deviation being around 50\% of the mean value) in ACE for \textit{ex vivo} bovine liver is attributed to the deviation from the Rayleigh distribution, rather to the biological variability \cite{Rubert2014}. A large variance due to spatial heterogeneity was also reported for a cervix remodelling assessment  study, where the difference in QUS parameters between early and late pregnant cervix were obscured by the intra-class variation \cite{McFarlin2015, guerrero2018}. The reliability of Controlled Attenuation Parameter (CAP, Fibroscan, Echosens, France), a promising commercial tool to detect and quantify hepatic steatosis, has been reported to be compromised due to the variability in attenuation estimates. For example, an IQR$>$40dBm was found to be indicative of poor diagnostic accuracy \cite{wong2017}.

Large axial window lengths and averaging over lateral and elevational RF lines reduce the spatial variation noise due to random scatterer positioning.  These approaches offer improved accuracy and precision, assuming uniform scattering within interrogated volume. Averaging also improves the power spectral estimation \cite{Liu2010}. Studies show that the Welch method, a refined periodogram that estimates the averaged periodogram of overlapped subwindows, was superior compared to rectangular, Hamming and Hanning windowing \cite{Liu2010}. While these techniques attain an improved estimation at the expense of spatial resolution, more recent works explore the possibility of extending the precision-resolution trade-off. Regularization incorporating a spatial prior was successful in simultaneously improving the resolution and the precision of QUS estimation for homogeneous regions \cite{coila2017, vajihi2018, rau2020}. However, uniform regularization, which was used in \cite{coila2017, vajihi2018, rau2020}, might  not be appropriate for biological tissue with the presence of inhomogeneities, as it will lead to oversmoothing in homogeneous regions in an attempt to compensate for the local inhomogeneities. In our previous work, we presented a proof-of-concept study based on a spatially weighted total variation regularization method, where the adaptive regularization parameter was a function of local inhomogeneity \cite{Deeba2019a}. The study was limited to the measurement of a single QUS parameter, ACE, in a 2D region-of-interest and was validated using a small \textit{in-vivo} liver study.

Another factor that might affect the clinical uptake of QUS imaging is the provision of visual guidance. CAP, which is performed in A-mode, lacks visual guidance, limiting ability to select the optimum region-of-interest. Several recent works on liver imaging have addressed the importance of B-mode and parametric image guidance to select an appropriate region-of-interest (ROI) that excludes vessels and extrahepatic areas \cite{fujiwara2018, Kanayama2013a}. 3D QUS imaging can further be advantageous over the 1-D or 2-D counterpart, allowing volumetric reconstruction, multimodal registration along with enhanced visualization and interpretation.

In this work, we present a multiparametric 3D weighted QUS (3D QUS) imaging system using a 3D total variation regularization method incorporating spatially variable regularization parameters. The proposed 3D adaptive total variation (TV) regularization exploits the spatial consistency information while preserving the variation in scattering properties in all three directions, thus would provide a rich spatial description of the tissue properties. The adaptive regularization is particularly important for QUS reconstruction in tissue regions with backscatter variation. We perform phantom validation, using phantoms with uniform and variable backscatter properties and demonstrate the superior QUS reconstruction performance of the proposed system compared to the baseline methods. Furthermore, the proposed method simultaneously generates multiple QUS maps, including ACE, IBC, and ESD, and therefore provides complementary information regarding the underlying tissue structure. The multiparametric approach could improve the class separability between normal and pathological tissue in the feature space. Toward this goal, we evaluate the benefit of a multiparametric approach over the single QUS counterpart in an \textit{in vivo} liver study including healthy volunteers and patients with hepatic steatosis. 

\section{Methodology}

\subsection{3D QUS: Proposed Volumetric Quantitative Ultrasound Parameter Estimation Algorithm}
In our previous work \cite{Deeba2019a}, we presented a single QUS parameter estimation method limited to a 2D region-of-interest. In the problem formulation, we utilized a simplified model for the backscatter coefficient, ignoring the frequency dependent term.  Herein, we introduce a 3D multiparametric QUS system, using an adaptive three-dimensional total variation regularization method (section \ref{method1}). As the adaptive regularization parameter, we utilize a 3D spatially weighted matrix as a function of envelope SNR (signal-to-noise ratio) deviation (section \ref{2.1.2}). We further adopt an improved model of the backscatter coefficient considering its frequency dependence. Using this model, we formulate  equations to extract scatterer statistics, such as effective scatterer diameter and effective acoustic concentration (section \ref{2.1.3}).
\subsubsection{Attenuation Coefficient Estimate and Integrated Backscatter Coefficient Measurement}
\label{method1}
A pulse-echo ultrasound system, with  the assumption of weak scattering (i.e. the Born approximation) and negligible effect of diffraction and refraction, can be modeled as a simple frequency domain multiplication of pulse characteristics and backscatter characteristics \cite{treece2005ultrasound}. The backscattered signal intensity, $S(f,z)$ obtained from a time-gated radio frequency signal window centered at depth $z$ from the transducer surface at frequency $f$ can be written as:

\begin{equation}
S(f,z) = G(f) A(f,z) B(f), 
\end{equation}
where $G(f)$ represents the transducer transfer function and transmit pulse. $A(f,z)$ represents the total attenuation effect from the transducer surface to the center of the respective time-gated window at depth $z$ and $B(f)$ is the backscatter term. 
For most soft tissue, $A(f,z)$ is approximately linearly proportional to the frequency \cite{wear2002gaussian}:
\begin{equation}
A (f, z) = e^{-4\alpha f z},
\end{equation}
where $\alpha$ (Np/cm/MHz) is the effective amplitude attenuation coefficient estimate for the round trip of the depth $z$. 

The backscatter term, $B(f)$, in its simplest form, can be expressed as a scalar parameter, as has been adopted in the scattering model used by Field II \cite{jensen1996field} and previous ACE computation works \cite{coila2017regularized, Deeba2019a}. However, a more appropriate model is required to capture the non-linear frequency dependence of the BSC term \cite{treece2005ultrasound, vajihi2018}:

\begin{equation}
B(f) = \beta f^n.
\end{equation}

  The reference phantom method \cite{yao1990backscatter}, a widely adopted method for QUS estimation, removes the system dependence (e.g. cancels out the system dependent term, $G(f)$) by normalizing the signal intensity backscattered from the tissue sample utilizing the one obtained from a well-characterized reference phantom under the same transducer and system settings. Therefore, the ratio of the intensity backscattered from the tissue sample to that from the reference at $(i,j,k)$ location at frequency $f_p$ can be written as:

  \begin{align}
  \label{eqn4}
  \begin{split}
RS(f,z)& = \frac{S_s(f,z)}{S_r(f,z)} = \frac{A_s(f,z) B_s(f)}{A_r(f,z)  B_r(f)}\\&= \frac{\beta_s f^{n_s}}{\beta_r f^{n_r}}e^{-4(\alpha_s-\alpha_r)f z},\\
\end{split}
  \end{align}
where $s$ and $r$ denote sample under experiment  and the reference, respectively.
Taking the natural logarithm of equation \ref{eqn4} leads to
\begin{align}
\label{eqn5}
\begin{split}
\ln RS(f,z) = -4(\alpha_s-\alpha_r)f z + \ln \frac{\beta_s}{\beta_r} + (n_s-n_r)\ln f
\end{split}
\end{align}
Substituting the following variables in equation \ref{eqn5} as
\begin{equation}
\label{eqn5a}
\ln RS(f,z) = y, \;  4fz = \phi, \;\alpha_r-\alpha_s = \alpha, \;  \ln \frac{\beta_s}{\beta_r} = \beta, n_s - n_r = n,
\end{equation}
we obtain,
\begin{equation}
\label{eqn6a}
y = \phi\alpha + \beta + n \ln f
\end{equation}
The system is defined in a $N_1 \times N_2 \times N_3 \;(N = N_1 N_2 N_3)$  spatial grid and a frequency band discretized at $M$ points is considered. We can represent the volumetric variables in vector format, for any variable $\textbf{a}\in \mathbb{R}^N$ with elements $a_{i,j,k}$ $\left( i\in[1,N_1],j\in[1,N_2],k\in[1,N_3]\right)$, reducing the forward model in equation \ref{eqn6a} to

\begin{equation}
\mathbf{y} = \mathbf{Ax} + \mathcal{N} (0, \sigma_N),
\end{equation}
where 
\[
\mathbf{A} = \begin{bmatrix}
     \boldsymbol{\phi}_1  & \mathbf{I} & \ln f_1 \mathbf{I}\\
     \boldsymbol{\phi}_2  & \mathbf{I} & \ln f_2 \mathbf{I}\\
     \vdots & \vdots & \vdots\\
      \boldsymbol{\phi}_M  & \mathbf{I}  & \ln f_M\mathbf{I} \\
   
    \end{bmatrix},
    \mathbf{y} = \begin{bmatrix} 
     \mathbf{y}_1\\
    \mathbf{y}_2\\
    \vdots \\ 
    \mathbf{y}_M\\
    \end{bmatrix},
    \mathbf{x} = \begin{bmatrix} 
     \boldsymbol{\alpha}\\
 \boldsymbol{\beta}\\
    \mathbf{n}\\
    \end{bmatrix}.
\]
Here, $\boldsymbol{\phi_m} \in \mathbb{R}^{N\times N}$ is a diagonal matrix, with the diagonal elements evaluated at propagation depth $\mathbf{z} \in \mathbb{R}^N$ for frequency $f_m \; (m\in[1,M])$ is given by

\begin{equation}
\boldsymbol{\phi_m}= 4f_m \textnormal{diag}(\mathbf{z}).
\end{equation} 
 The vector $\mathbf{y}\in\mathbb{R}^{NM}$ concatenates $M$ sub-vectors $\mathbf{y_m} \in \mathbb{R}^N$, with each $\mathbf{y_m}$ representing the volumetric power spectra ratio term evaluated at frequency $f_m$ and $\mathbf{I}\in \mathbb{R}^{N\times N}$ is a identity matrix. $\mathcal{N}(0,\sigma_N)$ is the Gaussian noise present in the measurement of $\textbf{y}$.

QUS parameters $\alpha$, $\beta$, and $n$ can abruptly change at the interface between two different tissue types, but these parameters will undergo gradual change (i.e. sparse spatial variation) within each tissue type. Therefore, with the piece-wise continuous assumption in all three spatial direction, we propose a 3D total variation regularization approach for the reconstruction of $\mathbf{x} = [\boldsymbol{\alpha, \beta, \mathbf{n}]}$ from the noisy estimation $\mathbf{y}$:

\begin{equation}
\label{eqn9}
\mathbf{\hat{x}} 
= \argmin_\mathbf{x} \norm{\mathbf{y}-\mathbf{Ax}}^2_2 +  \lambda_1|| {\boldsymbol{\alpha}}||_\mathbf{TV} + \lambda_2 ||\boldsymbol{\beta}||_\mathbf{TV} + \lambda_3 ||\boldsymbol{n}||_\mathbf{TV}  ,
\end{equation}
where the first term is the data fidelity term, the last three terms are the anisotropic 3D total variation regularization terms, and $\lambda_1$, $\lambda_2$ and $\lambda_3$ are the regularization weights. 

The 3D spatially weighted TV regularizer for a variable $\eta$ is defined as
\begin{align}
\label{eqn11}
\begin{split}
||\boldsymbol\eta||_{TV}  &= \sum_{i,j,k}W_{i,j,k}\left(\frac{1}{(h_{i+1,i})_{jk}}|{\eta_{i+1,j,k}-\eta_{i,j,k}}|\right.\\&+ \left. \frac{1}{(h_{j+1,j})_{ik}}| {\eta_{i,j+1,k}-\eta_{i,j,k}}|)+\frac{1}{(h_{k+1,k})_{ij}}| {\eta_{i,j,k+1}-\eta_{i,j,k}}|\right)
\end{split},
\end{align}
where the 3D weighted matrix $W_{i,j,k}$ is uniform for $\alpha$, and spatially variable for $\beta$ and $n$. $h_{i+1,i}, h_{j+1,j}, h_{k+1,k}$ are distances between the adjacent axial, lateral and elevational pixels respectively at corresponding spatial positions. The TV regularizer \cite{rudin1992nonlinear} minimizes the sum of the gradients by imposing sparsity promoting $l_1$- penalty in all three directions of the reconstructed QUS volume. By promoting signals with sparse gradients, TV minimization recovers piecewise-constant volumetric parameters.

Solving \eqref{eqn9} and \eqref{eqn5a} will give the values of volumetric $\boldsymbol\alpha_s$, $\boldsymbol\beta_s$ and $\boldsymbol n_s$. From the effective attenuation coefficient estimate, $\alpha$ obtained at depth $z$, the local attenuation coefficient estimate, $\alpha^{local}$ can be computed as: $$ \alpha_{i,j,k}^{local} = \frac{\alpha_{i,j,k}z_{i,j,k}-\alpha_{i-1,j,k}z_{i-1,j,k}}{z_{i,j,k}-z_{i-1,j,k}}.$$

The integrated backscatter coefficient (IBC) at $(i,j,k)$ location can be computed as:
$$
IBC =\frac{1}{f_M-f_1} \int_{f_1}^{f_M} {\beta f^{n}df}.
$$

 \begin{figure*}[]
    \centering

    \begin{subfigure}{.5\textwidth}
    \centering
        \includegraphics[height=.4\linewidth]{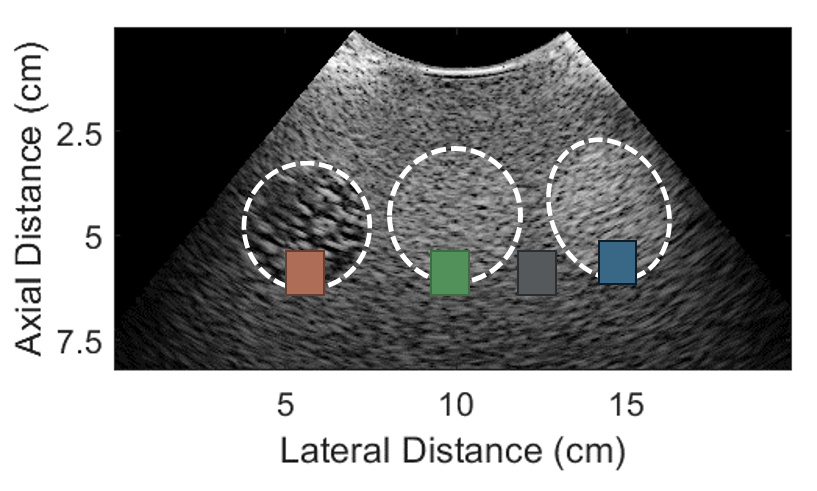}
            \caption{}
                   \label{fig:1a}
        \end{subfigure}%
        \begin{subfigure}{.5\textwidth}
           \centering
         \includegraphics[height=.4\linewidth]{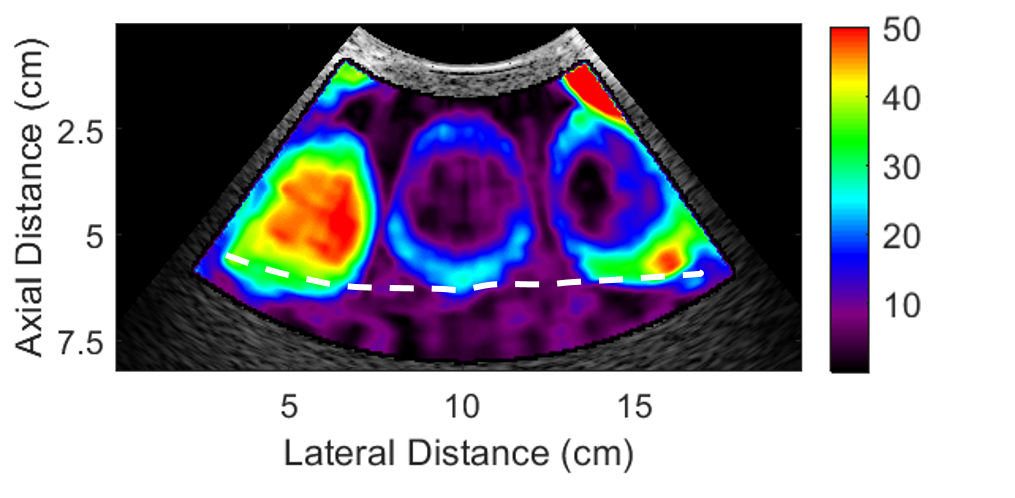}
        \caption{}
        \label{fig:1b}
    \end{subfigure}%
    \\
    \begin{subfigure}{.4\textwidth}
    \centering
        \includegraphics[height=.7\linewidth]{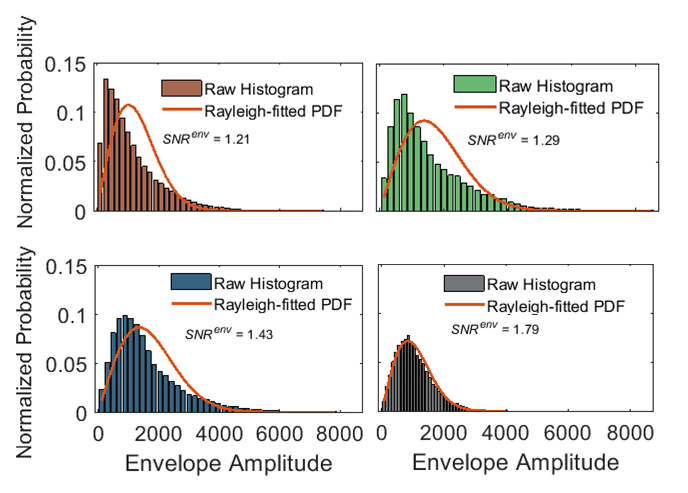}
              \caption{}
                         \label{fig:1c}
              \end{subfigure}%
                 \begin{subfigure}{.5\textwidth}
                 \centering
         \includegraphics[height=.45\linewidth]{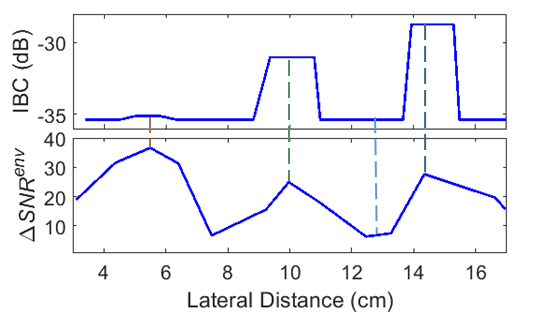}
         
        \caption{}
           \label{fig:1d}
    \end{subfigure}
    \caption{Efficacy of the adaptive regularization parameter $\Delta \mathit{SNR}^{env}$ to identify BSC variation within the ROI. (a) B-mode image with dashed white line outlining the three inclusions. Four color-coded ROIs are selected at the boundaries of the inclusions and the background. Inclusion 1 - red, inclusion 2 - green, inclusion 3 - blue, and background - gray.  (b) The adaptive regularization parameter $\Delta \mathit{SNR}^{env}$ map. (c) Rayleigh distribution fit associated to the selected ROIs. (d) IBC and $\Delta \mathit{SNR}^{env}$ along the curve shown in (b). The dashed lines correspond to the center point of the selected ROIs.} 
    \label{figure:1}
    \end{figure*}
\subsubsection{Selection of the Spatially Weighted Regularization Parameter}
\label{2.1.2}
Spectral-based methods, such as the reference phantom method, make the assumption that RF signal generation is a stationary scattering process, arising from a large number of randomly distributed scatterers identical in size. The RF envelope of such speckle pattern (namely fully developed speckle) can be modelled using the Rayleigh distribution \cite{goodman1985wiley}. There will be large measurement error and variance in QUS estimation when the  underlying assumption for Rayleigh distribution is violated. Such conditions occur for ROIs with varying backscatter properties or finite scatterer density ($<10$ scatterers per resolution cell) \cite{pawlicki2013, Rubert2014}. Therefore an adaptive regularization parameter is required, which is sensitive to the conditions leading to the underlying assumption for Rayleigh distribution.


Envelope SNR ($\mathit{SNR}^{env}$), defined as the ratio of the mean to the standard deviation of the RF signal envelope, is constant for a Rayleigh distribution and equals to $\mathit{SNR}^{opt}=$1.91. The $\mathit{SNR}^{env}$ value deviates from the Rayleigh level of 1.91 
for conditions involving finite scatterer density ($<10$ scatterers per resolution cell) or reduced degree of randomness (due to the presence of periodicity or clustering). Therefore, $\mathit{SNR}^{env}$ can be used to identify media with low scatterer density, a source of large variance in ACE measurement \cite{Rubert2014}.  Though $\mathit{SNR}^{env}$ cannot discriminate media with different backscatter characteristic for large scatterer densities ($>10$ scatterers per resolution cell), it can identify the boundary of two such media, as the Rayleigh assumptions of identical diameter and random distribution do not hold at the boundary.  

We define a normalized $\mathit{SNR}^{env}$ parameter, envelope SNR deviation ($\Delta \mathit{SNR}^{env}$) \cite{Deeba2019}, given at a spatial location $(i,j,k)$ by
 \begin{equation}
     \Delta {\mathit{SNR}}^{env}_{i,j,k}\  = \frac{|{\mathit{SNR}}^{env}_{i,j,k} -\mathit{SNR}^{opt}|}{\mathit{SNR}^{opt}} \times 100\%.
 \end{equation}
 Previous works qualitatively demonstrated the association between $\Delta {\mathit{SNR}}^{env}$ and backscatter variation from the visual inspection of inhomogeneity, lacking the quantitative description of BSC parameters \cite{Deeba2019, Deeba2019a}. In this work, we investigate the dependence of $\Delta {\mathit{SNR}}^{env}$ on the quantitative backscatter parameters (i.e. IBC) and therefore validate the efficacy of $\Delta {\mathit{SNR}}^{env}$ to detect the backscatter variation within the Region-of-interest (ROI). 

We perform a feasibility analysis on RF data acquired from a custom built ultrasound phantom (Table \ref{tab1}) with three inclusions of similar ACE and different backscatter characteristic, manufactured by CIRS (Norfolk, VA, USA). Inclusion 1 (with low scatterer density) can be identified from the background with significantly high $\Delta \mathit{SNR}^{env}$, whereas inclusion 2 and inclusion 3 have similar $\Delta \mathit{SNR}^{env}$ as the background (Fig.\ref{fig:1b}). However, the boundaries for all three inclusions are highlighted with increased $\Delta \mathit{SNR}^{env}$. This trend is also evident in the Rayleigh distribution fit to the RF envelope amplitude for selected ROIs, where the envelope amplitude in the ROIs on the boundaries deviate from the Rayleigh distribution (Fig. \ref{fig:1c}). The spatial variation in $\Delta \mathit{SNR}^{env}$ correlates with the change in decibel scaled IBC (dB) (Fig. \ref{fig:1d}). Therefore, $\Delta \mathit{SNR}^{env}$ is a suitable regularization parameter which can identify backscatter variation as well as sufficiently low scatterer density.

We propose a 3D spatially weighted matrix, $\{W_{i,j,k}\}\in \mathbb{R}^N$ as a function of $\Delta \mathit{SNR}^{env}$ to adaptively regularize the backscatter parameters, $n$ and $\beta$:

\begin{equation}W_{i,j,k}\left( {\Delta SN{R^{env}_{i,j,k}}} \right) = \frac{a}{{1 + \exp \left[ {b.\left( {\Delta SN{R^{env}_{i,j,k}} - \Delta \mathit{SNR}^{env}_{\min }} \right)} \right]}},\end{equation}
where $a$ and $b$ are constants and $\Delta \mathit{SNR}^{env}_{\min}$ is a nominal $\Delta \mathit{SNR}^{env}$ value for homogeneous region. For $\Delta \mathit{SNR}^{env}<<  \Delta \mathit{SNR}^{env}_{\min}$, the weighting has little effect on the regularization. As the weight decreases with the increase in $\Delta \mathit{SNR}^{env}$, the regularization effect on the backscatter terms is relaxed.

\begin{figure}
    \centering
    \begin{subfigure}{.35\textwidth}
    \centering
        \includegraphics[trim={5cm 19.5cm 5cm 0cm},clip,scale=0.45]{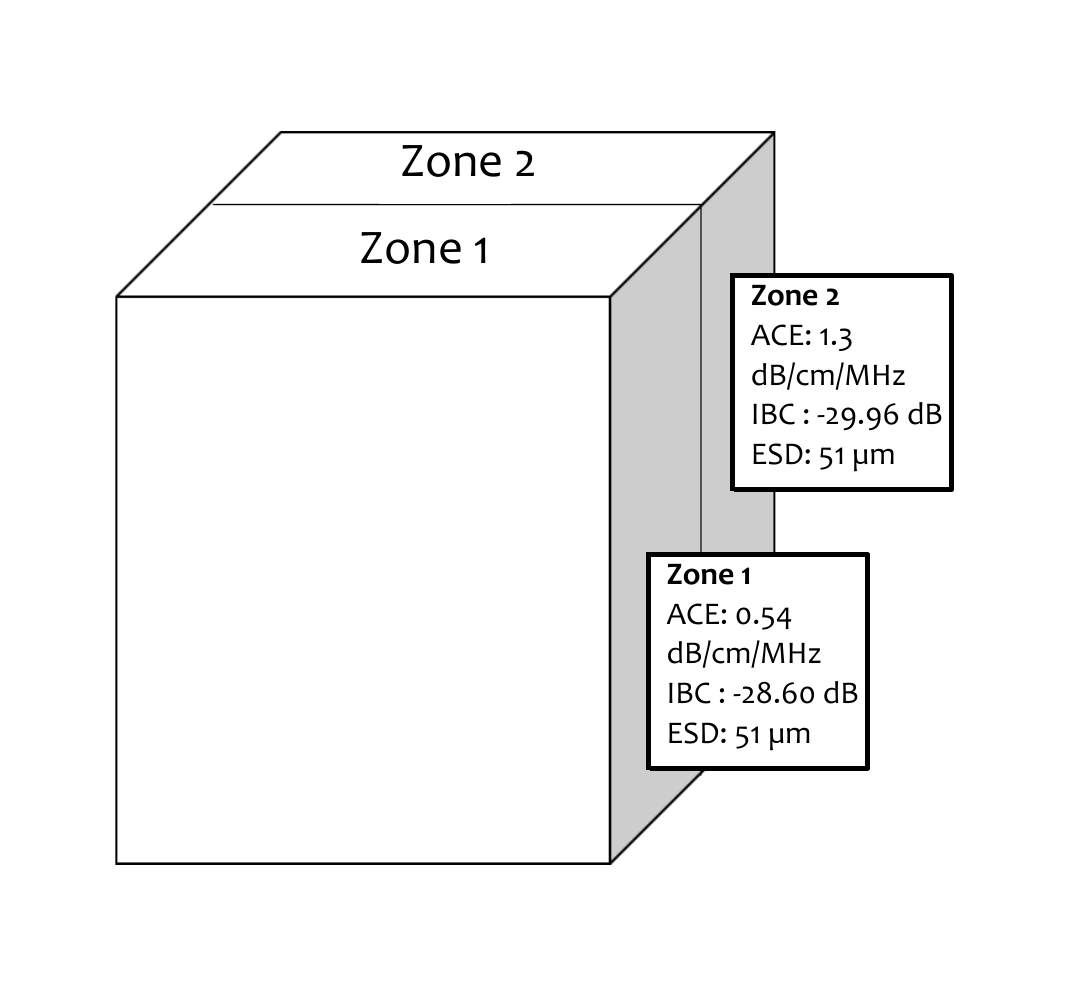}
            \caption{}
                   \label{fig:2a}
        \end{subfigure}%
        \\
        \begin{subfigure}{.35\textwidth}
           \centering
         \includegraphics[trim={3cm 20cm 5cm 2cm},clip,scale=0.55]{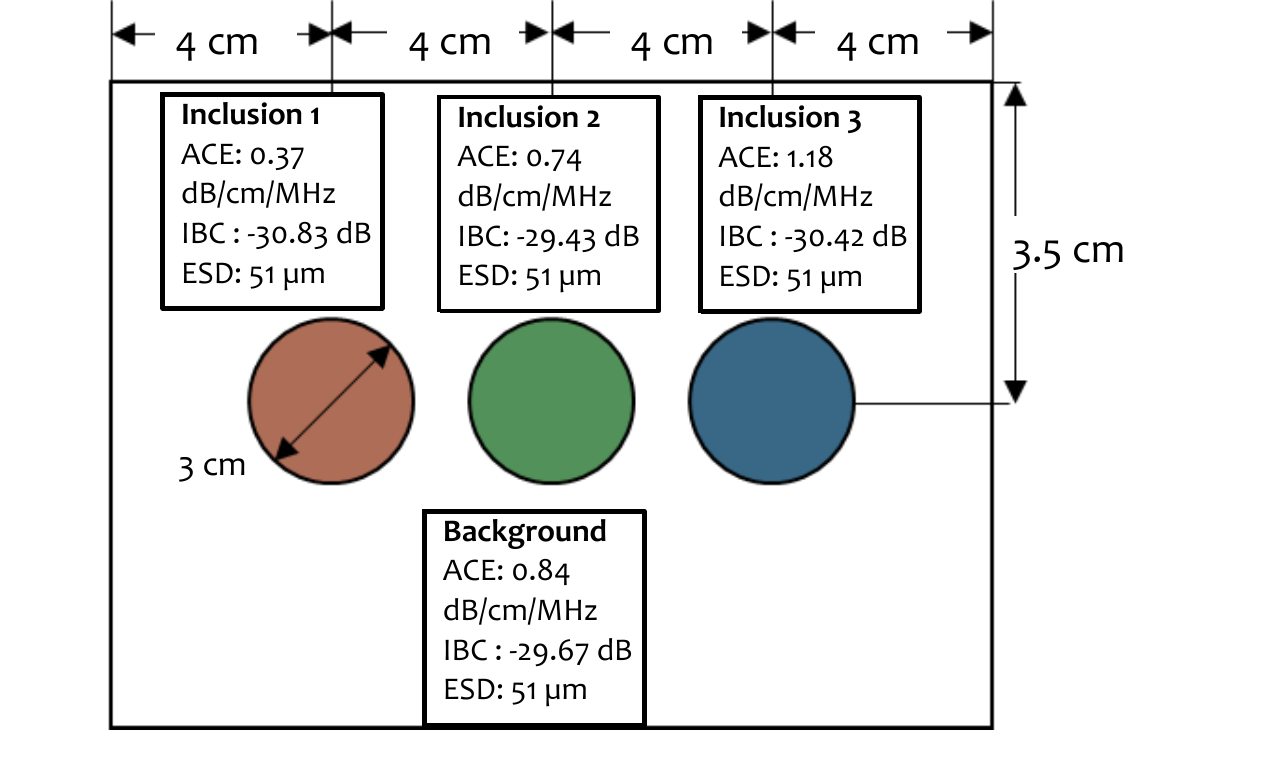}
        \caption{}
        \label{fig:2b}
    \end{subfigure}%
    \\
     \begin{subfigure}{.35\textwidth}
           \centering
         \includegraphics[trim={2.5cm 20cm 5cm 2cm},clip,scale=0.55]{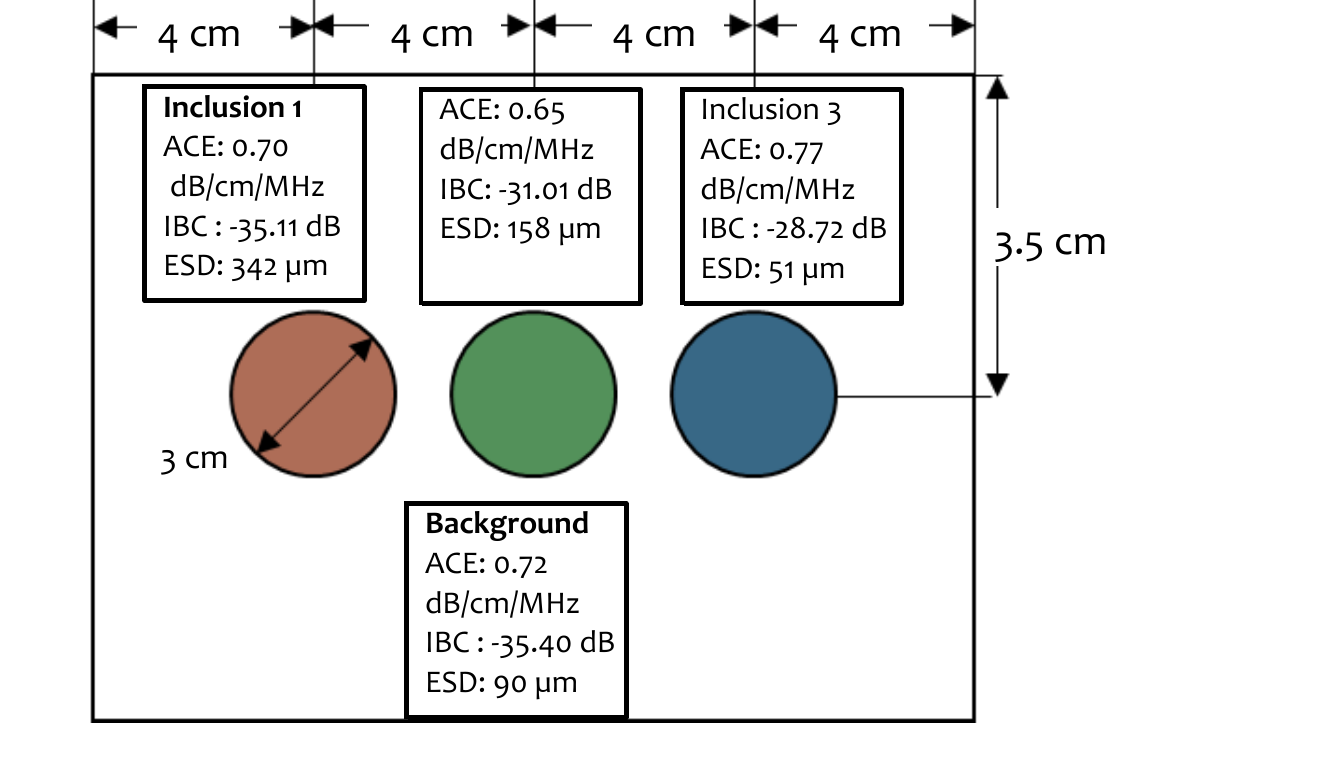}
        \caption{}
        \label{fig:2b}
    \end{subfigure}%
    \caption{Diagram of the custom-made CIRS phantoms.(a) Phantom 1 with two vertical layers; and (b) phantom 2 with three spherical inclusions (variable ACE and similar backscatter (IBC) and scatterer diameter (ESD) property) and  (c) phantom 3 with three spherical inclusions (similar ACE and variable backscatter (IBC) and scatterer diameter (ESD) property)} 
    \label{figure:2}
    \end{figure}
\subsubsection{Effective Scatterer Diameter (ESD) Measurement}{\label{2.1.3}}
The backscatter coefficient, when expressed in terms of an acoustic intensity form factor, can be used to extract the effective scatterer diameter (ESD) and effective acoustic concentration (EAC). We utilize a Gaussian form factor $F_{gauss} = \exp(-0.827k^2 a_{eff}^2) $ \cite{insana1990describing}, which assumes scattering sources with continuously varying acoustic properties randomly suspended in a fluid-like media. The backscatter coefficient, $B$ can be written as:

\begin{equation}
\label{6}
B(f) = \frac{1}{9} k^4 a_{eff}^6 n\bar{  \gamma}_0^2F_{gauss}.
\end{equation} 
Here, $k = 2\pi f/c$ is the wave number, $c$ is the sound speed, and $a_{eff}$ is the effective scatterer radius. $n\bar\gamma_0^2$ is the net scattering strength, which is the product of the concentration of scatterers $n$ times the relative impedance difference between the scatterers and surrounding tissues, $\bar{\gamma}_0^2$. $B(f)$ $(= \beta f^n)$ is the backscatter coefficient measured using the 3D adaptive regularization method (section \ref{method1}).

Dividing by $f^4$ and then taking the natural logarithm on both sides of \eqref{6} we get:
\begin{equation}
\ln{B_s(f)\cdot f^{-4}} = \ln\left(\frac{1}{9}\left({\frac{2\pi}{c}}\right)^4  a_{eff}^6 n \bar\gamma_0^2\right)-0.827\left(\frac{2\pi f}{c} \right)^2 a_{eff}^2
\end{equation}
We can estimate $a_{eff}$ and $n\bar\gamma_0^2$ by fitting a regression line, $y= mx +C$, where $y = \ln{B_s(f)\cdot f^{-4}}  $, $x = f^2$, and 
\begin{equation}\label{m}
m = -0.827\left(\frac{2\pi}{c} \right)^2 a_{eff}^2, 
\end{equation}
\begin{equation}\label{c}
C =  \ln\left(\frac{1}{9}\left({\frac{2\pi}{c}}\right)^4  a_{eff}^6 n \bar\gamma_0^2\right).
\end{equation}

The effective scatterer diameter ($ESD = 2a_{eff}$) can be estimated from \eqref{m} and effective acoustic concentration ($EAC = 10 \log(n \bar \gamma_0^2)$) can be obtained solving \eqref{m} and \eqref{c} \cite{rohrbach2018high}. The `Model Inversion' technique \cite{oelze2002characterization} of measuring scattering properties by comparing measured backscatter spectrum to the theoretical one and fitting a regression line has been utilized in several previous works \cite{rohrbach2018high, nizam2020classification}. This technique is computationally efficient over the traditional minimum average squared deviation (MASD) technique \cite{insana1990describing}. However, both of these techniques employ single time-gated window-based backscatter coefficient measures, which are prone to large measurement error. A previous work showed an improvement in QUS estimation using a nearest neighborhood averaging, which exploits the similarity among QUS measures of neighbouring windows \cite{nizam2020classification}. In the proposed method, ESD is derived from a TV regularized backscatter measure and therefore exploits the sparse spatial variation of QUS parameters. 
\begin{table*}[!t]
\caption{\label{tab1}Ground truth values for the phantoms}
\centering
\begin{tabular}{|p{2.25cm}|p{2cm}|l|p{4cm}|p{3cm}|}
\hline
\multicolumn{2}{|c|}{Phantom}  & ACE (dB/cm/MHz) & IBC (dB) & ESD ($\mu m$)
\\
\hline
\multirow{2}{*}{Phantom 1}
			 & Zone 1 & 0.54 & -28.60 & 51\\\cline{2-5} 
             & Zone 2 & 1.3& -29.96 & 51
\\\hline   

\multirow{4}{*}{Phantom 2} 
			& Background & 0.84 & -29.67 & 51\\\cline{2-5} 
			& Inclusion 1 & 0.37 & -30.83 & 51 \\\cline{2-5} 
			& Inclusion 2 & 0.74  & -29.43 & 51 \\\cline{2-5} 
			& Inclusion 3	& 1.18	& -30.42 & 51\\\cline{2-5} 
\hline
\multirow{4}{*}{Phantom 3} 
			& Background  & 0.72 &  -35.40 & 90\\\cline{2-5}
			& Inclusion 1	& 0.70	&  -35.11 & 342\\\cline{2-5}
			& Inclusion 2 & 0.65 & -31.01 & 158\\\cline{2-5}
			& Inclusion 3 & 0.77 &  -28.72 & 51
\\\hline
\end{tabular}
\end{table*}

\subsection{Data Sets}
\subsubsection{Phantoms}
The system was validated on three custom-built phantoms manufactured by CIRS (Norfolk, VA, USA). All three phantoms consisted of proprietary Zerdine hydrogel polymer and glass bead scatterers (Potter Industries, Malvern, PA, USA). Phantom 1 (as sketched in Fig. \ref{fig:2a}) has two vertical homogeneous zones with similar backscatter properties and variable ACE. Zone 1 of phantom 1 is used as the reference phantom. Phantom 2 and phantom 3 (sketched in Fig. \ref{fig:2b}) both include three spherical inclusions. The acoustic properties within the inclusion and the background were adjusted by adding soda lime glass microspheres (Spheriglass®, Potters Industries Inc., PA, USA) of varying diameter and concentration. The inclusions in phantom 2 have variable ACE properties and similar backscatter and scatterer size properties, whereas the inclusions in phantom 3 have similar ACE with variable backscatter and scatterer size properties. The ground truth values for the QUS parameters reported by CIRS have been listed in Table \ref{tab1}. For all the experiments, zone 1 of phantom 1 was used as the reference phantom.

\begin{figure*}[]
        \includegraphics[trim={0 18.5cm 0cm 0cm},clip,scale=0.5]{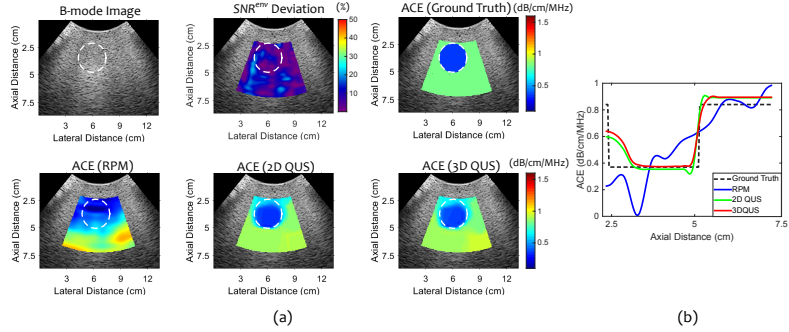}

    \caption{(a)  $\Delta \mathit{SNR}^{env}$ map, ground truth ACE map and reconstructed ACE map using different methods for inclusion 1 in phantom 2 (variable ACE and approximately uniform backscatter and scatterer size property) and (b) the profile of the reconstructed ACE along the axially central line through the inclusion.} 
    \label{figure:5}
    \end{figure*}
 
\subsubsection{\textit{In vivo} Liver}    
A cross-sectional study was performed, which included a cohort of twenty-one adult participants, with 13 healthy controls (defined as MRI-PDFF$<$5\%) and 8 NAFLD patients (MRI-PDFF$>$5\%). This study ( H14-01964) was performed under written informed consent after approval by the UBC Clinical Research Ethics Board. All the participants underwent MRI and ultrasound examination between May 2017 and May 2018. MRI was performed on a 3.0 T system (Philips Achieva, Philips Medical Systems) with a double double echo (DDE) sequence, consisting of two dual fast field echo sequences. A three-point Dixon method was used to determine the MRI-PDFF (proton density fat fraction). MRI-PDFF computed from the MRI data has been found to be strongly correlated with histological steatosis grading \cite{noureddin2013utility}. Therefore, MRI-PDFF was used as a gold standard for hepatic steatosis detection, where an MRI-PDFF$>5\%$ was defined as presence of hepatic steatosis. The ultrasound data were acquired from the right lateral intercostal  window of the volunteer, laid in a supine position on a shaker board propagating the mechanical excitation. 
\subsection{Data Acquisition and Processing}
\subsubsection{3D QUS System Overview}
\label{2.2.1}

The proposed volumetric QUS imaging technique was implemented on a standard 3D ultrasound system, that was also used for Shear Wave Absolute Vibro-Elastography (S-WAVE) developed in our lab \cite{abeysekera2016}. The addition of steady state shear waves for S-WAVE has negligible effect (shown later) on the QUS measures so the data acquisition for QUS can be performed either simultaneously with S-WAVE or serially, i.e. after removing shear wave excitation. The 3D US system, used in this work, comprises an Ultrasonix SonixTouch ultrasound machine (Analogic, Richmond, BC, Canada), a m4DC7-3/40 4D transducer, and a linear voice coil actuator for shear wave motion generation. The transducer center transmit frequency was set to 3.33 MHz whereas the imaging depth was set to 15 cm with a focus at 13 cm.  The customized motor control module allows the acquisition of volumetric RF data consisting of 15 planes spaced at {$1^\circ$ angular distance, with 25 frames acquired at each plane location at an effective frame rate of 250 frame-per-second (fps). Each 2-D frame in the 3-D volume consists of 64 RF scanlines covering a lateral field-of-view of $79^\circ$. The sampling rate was 20 MHz, resulting in 3870 samples in each scanline.

\subsubsection{RF data processing and analysis}
\label{2.2.2}
The proposed method was implemented in MATLAB 2018a (The MathWorks Inc., Natick, MA, USA). The optimization problem was solved using the convex optimization toolbox CVX in MATLAB \cite{grant2009cvx}. We compare the performance of the proposed 3D QUS method against the baseline reference phantom method (RPM) and a uniform 2D QUS method utilizing 2D uniform total variation regularization \cite{coila2017}, applied on each of the elevational planes separately.

We manually selected ROIs within the tissue or phantom, excluding the near-field data suffering from ringdown and other artifacts associated with transducer-to-phantom interface. Each ROI included 25 temporal frames. Each frame within the ROIs were divided into time-gated windows with $80\%$ overlap in both lateral and axial direction. 
The optimum dimensions for the time-gated windows for RPM were found to be 20 scan lines laterally and 50 wavelength ($\lambda$) (600 samples or 2.3 cm) axially. Here, the axial extent of one wavelength is approximately 0.46 mm. On the other hand, the dimensions for the time-gated windows for the regularization methods were selected as 5 scanlines laterally and 1.2 cm (300 samples or $25 \lambda$) axially. The Welch method \cite{welch1967use} was used to obtain the power spectrum from the RF scan lines within each time-gated window. We considered the $-20$ dB bandwidth of the received power spectrum as the usable frequency range.  

The power spectra obtained from a single realization suffers from large variance due to statistical fluctuations arising from a random media. Compounding $n$ statistically independent but equivalent realizations have been found to improve the power spectral estimate by reducing the standard deviation proportional to $1/\sqrt{n}$. Previously, different spatial and angular compounding based on spatial translations or rotations have been adopted to decorrelate the signal allowing independent realizations of RF data \cite{li1994evaluational, gerig2004correlation, herd2011improving}. In the proposed method, we perform averaging of the periodograms spatially and temporally, using 5 lateral scanlines and 25 temporal frames, to compute the power spectral density. To further improve the QUS estimation using RPM, we average the periodograms over the elevational planes resulting in a single QUS plane. The 2D QUS method applies 2D uniform regularization on each of the elevational planes separately, whereas the 3D QUS method applies 3D adaptive regularization using all the elevational planes.

We tune the regularization weights $\lambda_1$, $\lambda_2$ and $\lambda_3$ to obtain optimum QUS reconstruction results for the homogeneous phantom (phantom 1) and the variable ACE phantom (phantom 2). The regularization weights used for all QUS results presented in this paper using the 2D QUS method were $\lambda_1 =2^0, \lambda_2 = 2^{-2}, \lambda_3 = 2^{-3}$, whereas $\lambda_1 =2^{0.5}, \lambda_2 = 2^{0}, \lambda_3 = 2^{-0.5}$ were used for the 3D QUS method.

\subsection{Evaluation Metrics}
The mean and standard deviation of ACE, IBC and ESD were measured and reported. The accuracy and precision were assessed using the mean absolute percentage error and standard deviation of absolute percentage error, respectively. Absolute percentage error of a quantity $\textbf{a}\in \mathbb{R}^N$, is defined as:
\begin{equation}
APE = \frac{\mid \textbf{a}_{estimated}-\textbf{a}_{actual}\mid}{\textbf{a}_{actual}}\times 100\%,
\end{equation}
where $APE$ is the absolute percentage error, $\textbf{a}_{estimated}$ and $\textbf{a}_{actual}$ represent the reconstructed and the ground truth measure of a QUS quantity, respectively.

The detectability of an inclusion in a QUS map was evaluated using contrast-to-noise ratio (CNR) ratio:

\begin{equation}
CNR = \frac{\mid\mu_{inc}-\mu_{bg}\mid}{\sqrt{\sigma_{inc}^2+\sigma_{bg}^2}},
\end{equation} 
where $\mu$ and $\sigma$ are the mean and standard deviation of a QUS quantity within the inclusion and the background denoted by the subscripts $inc$ and $bg$, respectively. The inclusions were delineated from the B-mode images, refined using the position and dimension information provided by the manufacturer.

To evaluate steatosis detection performance for \textit{in vivo} liver, the correlation between MRI-PDFF and QUS measures were calculated using Spearman's rank correlation coefficient. Also, the classification accuracy and the area under the ROC curve (AUROC) were measured to evaluate the classification performance using different combination of QUS parameters.

\section{Experiments and Results}
\subsection{Phantom Validation}\label{3.1}

The ACE reconstruction results for phantom 2 (variable ACE and similar backscatter and scatterer size properties) are shown in Fig. \ref{figure:5}. From the $\Delta \mathit{SNR}^{env}$ map, it is evident that the variation in ACE does not cause deviation of $\mathit{SNR}^{env}$ values from $\mathit{SNR}^{opt}$. With $\Delta \mathit{SNR}^{env}<\Delta \mathit{SNR}^{env}_{min}$, the spatial weighting has little effect on the regularization. Therefore, QUS reconstruction results from the 2D QUS method with uniform regularization and the 3D QUS with adaptive regularization are similar, where both methods clearly identify the low attenuation inclusion from the high attenuation background. The RPM, on the other hand, detects the variation in ACE with a poorer resolution compared to the regularization methods.

Figure \ref{figure:6} shows the results for phantom 3 (approximately uniform ACE and variable backscatter and scatterer size properties). As previously discussed in section \ref{2.1.2}, the variation in backscatter characteristic is associated with a variation in $\Delta \mathit{SNR}^{env}$ map. The 3D QUS method, that incorporates the $\Delta \mathit{SNR}^{env}$ information for enabling adaptive regularization, yields more accurate reconstruction of the the IBC map compared to its uniform regularization counterpart, 2D QUS (Fig. \ref{figure:6}(a)). The RPM failed to generate a reasonable IBC result. The ACE maps obtained from the three methods have also been shown in Fig. \ref{figure:6} (c). Both the RPM and the 2D QUS method exhibit consecutive undershoot and overshoot in ACE measurement at the boundary between medium with different backscatter properties, a trend described in previous literature  \cite{pawlicki2013, Deeba2019a}. The proposed 3D QUS method, on the contrary, was able to overcome the issue, yielding accurate ACE results.

\begin{figure*}[]
        \includegraphics[trim={0 14cm 0 0cm},clip,scale=0.5]{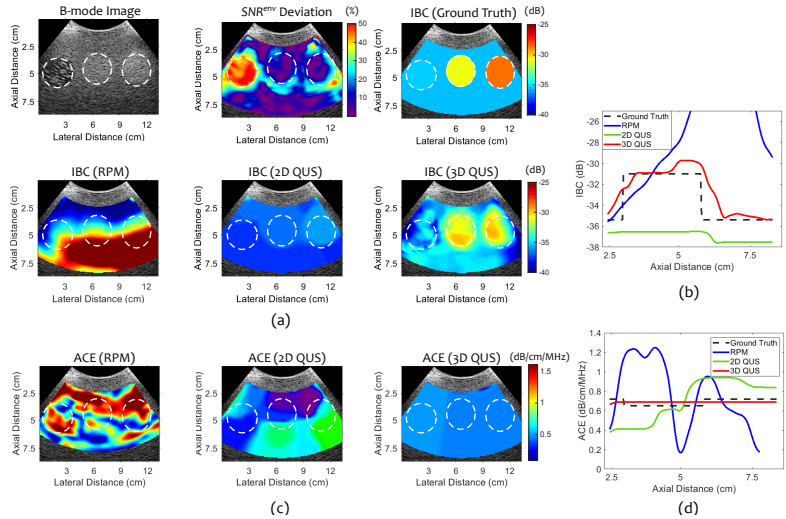}

    \caption{(a) $\Delta \mathit{SNR}^{env}$ map, ground truth IBC map and reconstructed IBC maps using different methods for phantom 3 (similar ACE and variable backscatter and scatterer size property). Inclusion 1, 2, and 3 have been annotated in all the maps. (b) The profile of the reconstructed IBC along the axial central line through inclusion 2. (c) Reconstructed ACE maps using different methods for phantom 3 and (d) the profile of the reconstructed ACE along the axial central line through inclusion 2.}
    \label{figure:6}
    \end{figure*}
The ESD map for phantom 3 has been shown in Fig. \ref{figure:7}. The RPM and the 2D QUS method using uniform regularization were not successful to reconstruct the ESD map, as were expected from the inaccurate estimation of IBC map. The proposed 3D QUS method results in improved ESD reconstruction.

\begin{table*}[!t]
\caption{\label{tab2} Estimated mean and standard deviation of absolute percentage error for the reconstructed ACE and IBC maps for the experimental phantoms.}
\centering
\begin{tabular}{|c|c|c|c|c|c|c|}
\hline
\multirow{3}{*}{Phantom} & \multicolumn{6}{|c|} {ACE (dB/cm/MHz)} 
\\
\cline{2-7} 
 & \multicolumn{3}{|c|} {Mean absolute percentage error (\%)} & \multicolumn{3}{|c|} {Std. of absolute percentage error (\%)} 
 \\
\cline{2-7} 
 & RPM & 2D QUS & 3D QUS &  RPM & 2D QUS & 3D QUS 
\\
\cline{1-7} 
1 & 7.43 & 2.97 & {2.96} &  4.61 & {0.35} & 0.71 
\\
\cline{1-7} 
 2 & 23.05 & {8.23} & 9.68 &  16.75 & 6.77 & {6.06} 
\\
\cline{1-7} 
3 & 61.15 & 19.93 & {7.34} &  37.52 & 8.90 & {2.3e-6} 
\\
\hline
\multirow{3}{*}{Phantom} & \multicolumn{6}{|c|} {IBC (dB)} 
\\
\cline{2-7} 
 & \multicolumn{3}{|c|} {Mean absolute percentage error (\%)} & \multicolumn{3}{|c|} {Std. of absolute percentage error (\%)} 
 \\
\cline{2-7} 
 & RPM & 2D QUS & 3D QUS &  RPM & 2D QUS & 3D QUS 
\\
\cline{1-7} 
1 & {6.30} & 24.94 & 23.29 &  3.55 & {6.2e-7} & 0.37
\\
\cline{1-7}
2 & 135.97 & 34.15 & {24.15} &  125.17 & {1.91} & 10.66
\\
\cline{1-7}
3 & 2247.2 & 45.06 & {22.83} & 
 866.25 & {0.56} & 10.48
 \\
 \cline{1-7}

\end{tabular}
\end{table*}

 \begin{figure*}[]
        \includegraphics[trim={0 19cm 0 0cm},clip,scale=0.5]{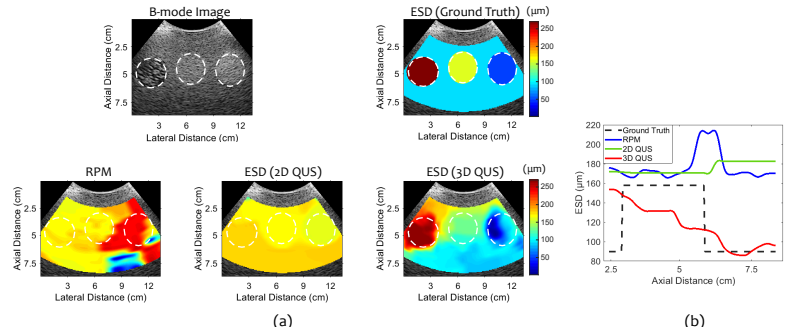}

    \caption{(a) $\Delta \mathit{SNR}^{env}$ map and reconstructed ESD maps using different methods for phantom 3 (approximately uniform ACE and variable backscatter and scatterer size property) with inclusion 1, 2, and 3. (b) The profile of the reconstructed ESD along the axial central line through inclusion 2.}
    \label{figure:7}
    \end{figure*}

For the quantitative comparison of the QUS parameters, we compare the mean and standard deviation of the ACE in phantom 2 and IBC and ESD in phantom 3 for the inclusions and the background. For IBC, we compute the standard deviation of the IBC error expressed in dB. The results are shown in Fig. \ref{figure:8}. We also report the mean absolute percentage error and standard deviation of absolute percentage error results in Table \ref{tab2} and \ref{tab3}. In case of ACE, both the 2D QUS method and the 3D QUS method reach close to the ground truth value with high precision, enabling accurate ACE characterization. However, for IBC and ESD estimation, the proposed 3D QUS method consistently outperforms both the other methods. 
 \begin{figure*}[h]
    \centering
        \includegraphics[width=.3\linewidth]{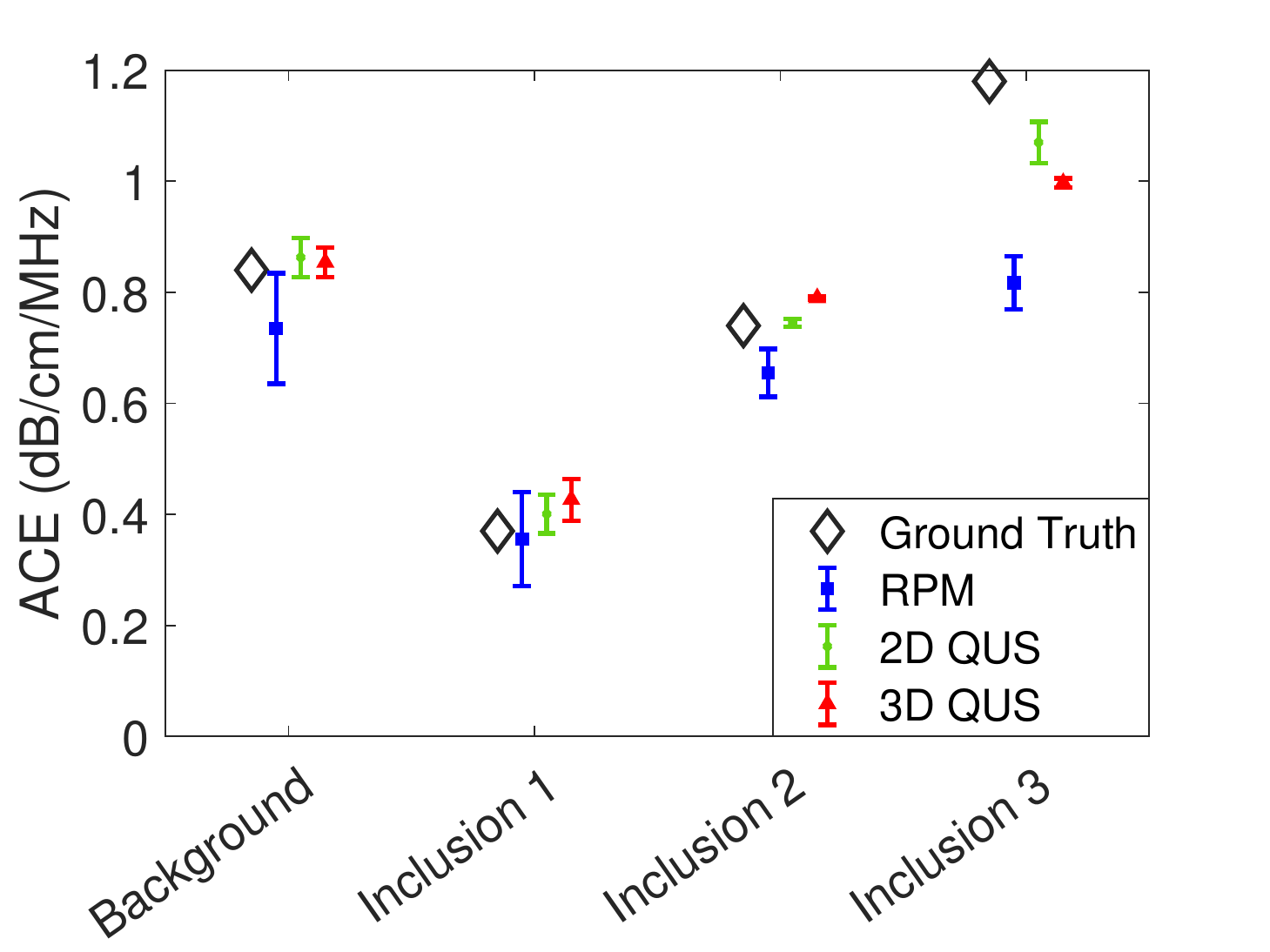}
         \includegraphics[width=.3\linewidth]{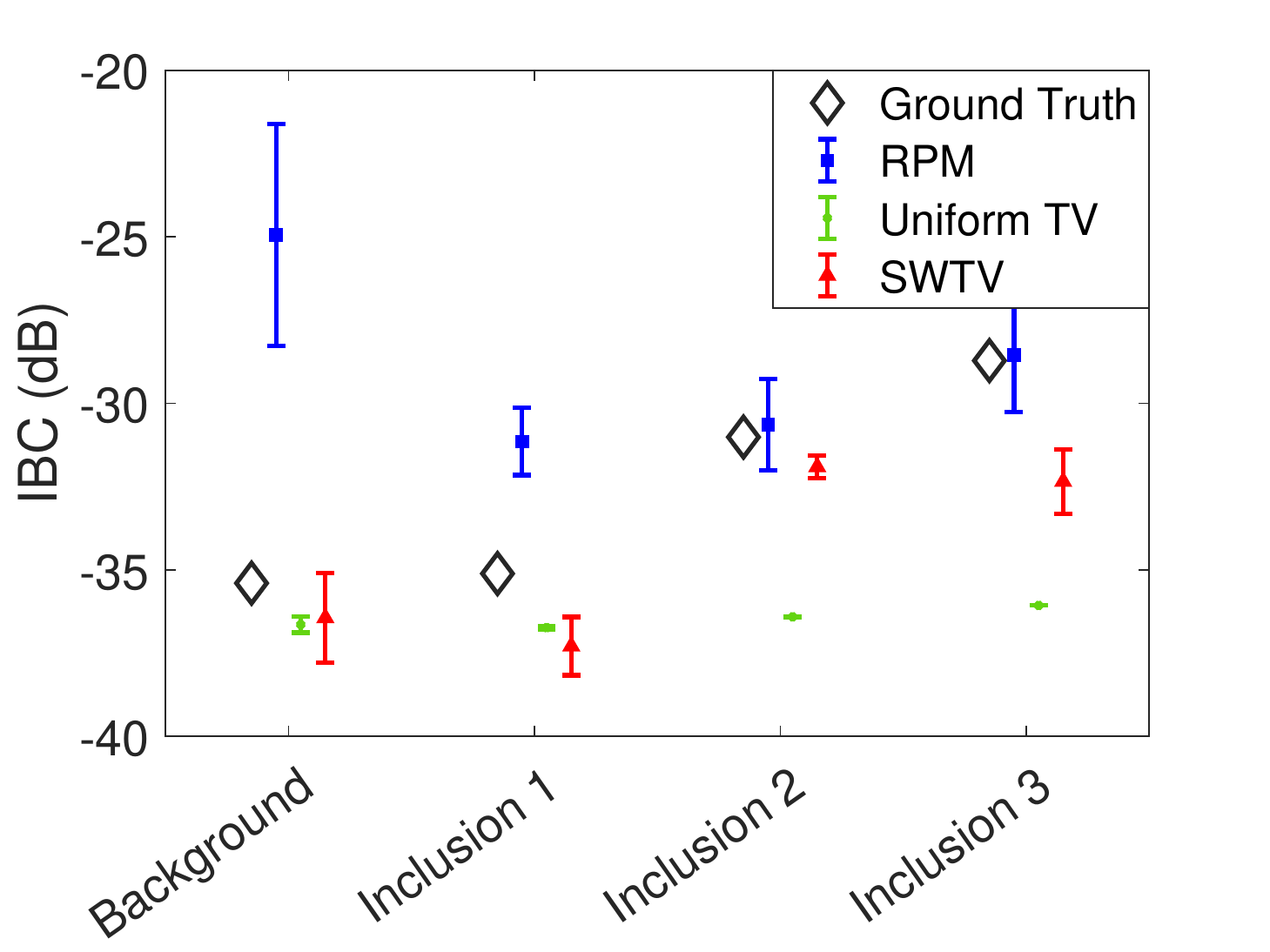}
        \includegraphics[width=.3\linewidth]{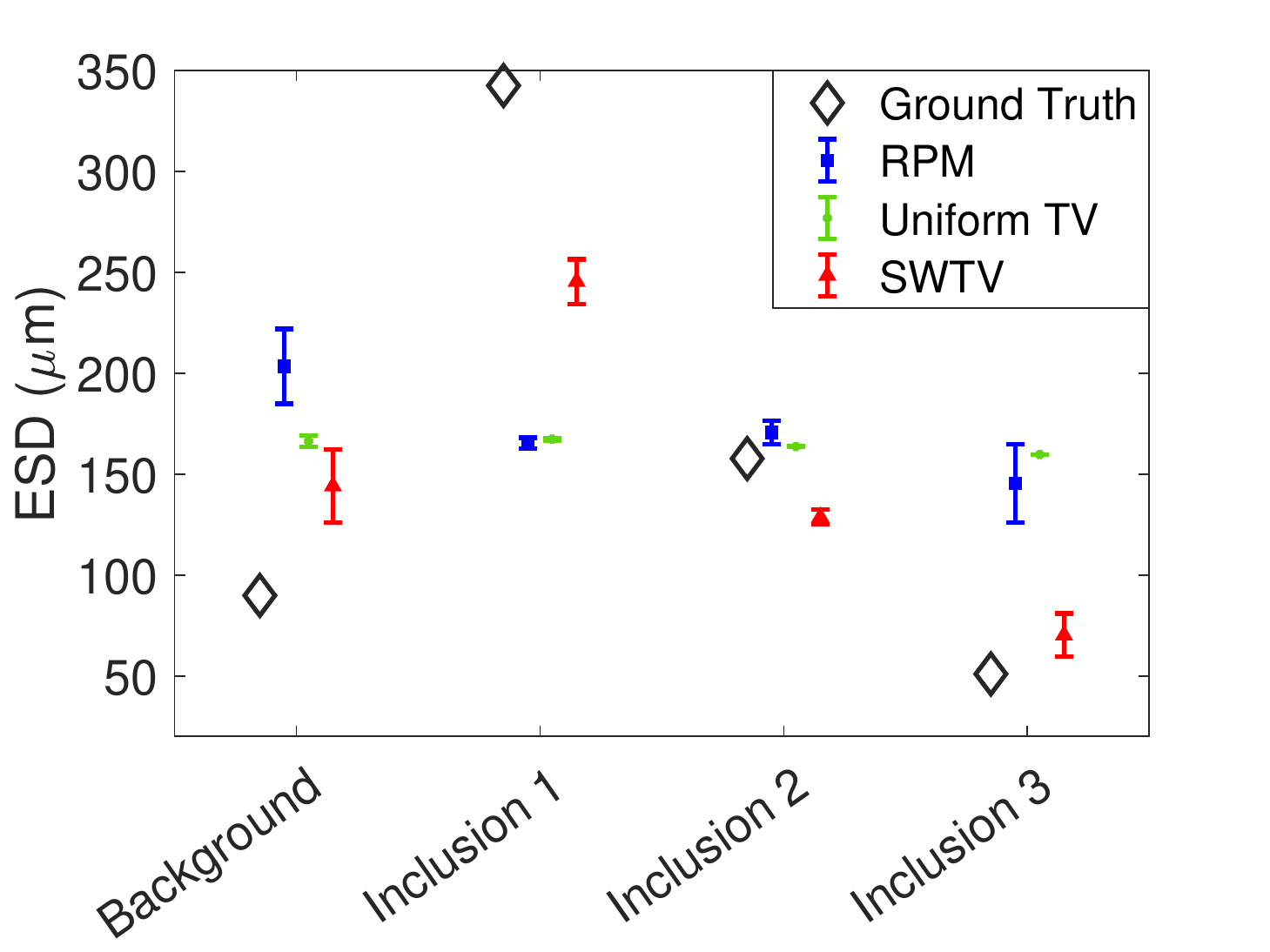}       
    \caption{Mean and standard deviation of ACE in phantom 2 (left), IBC (middle) and ESD (right) in phantom 3 using different methods.} 
    \label{figure:8}
    \end{figure*}
    
  We evaluate the CNR of the QUS parameters for different methods. Figure \ref{figure:9} shows the mean and standard deviation computed over the three inclusions in respective phantom. The mean CNR values obtained using the proposed 3D QUS method were larger than the other two methods. Compared to the baseline RPM, the 3D QUS method results in around 4.2 fold, 1.8 fold and 2.9 fold increase in the CNR for ACE, IBC, and ESD, respectively. Also, the 3D QUS improves the CNR by 1.4 times and 1.9 times compared to the 2D QUS method (uniform regularization) for IBC and ESD, respectively.  
    \begin{table*}[!t]
\caption{\label{tab3} Estimated mean and standard deviation of absolute percentage error for the reconstructed ESD map for phantom 3.}
\centering
\begin{tabular}{|c|c|c|c|c|c|c|}
\hline
 \multirow{2}{*}{Region}  & \multicolumn{3}{|c|} {mean absolute percentage error (\%)} & \multicolumn{3}{|c||} {Standard deviation of absolute percentage error (\%)}
 \\
\cline{2-7} 
 & RPM & 2D QUS & 3D QUS &  RPM & 2D QUS & 3D QUS 
\\
\cline{1-7} 
Inclusion 1 & 51.71 & 51.14 & 28.35 &  1.56 & 0.34 & 6.44 
\\
\cline{1-7} 
 Inclusion 2 & 8.11 & 3.78 & 18.36 &  7.36 & 0.07 & 4.78
\\
\cline{1-7} 
 Inclusion 3 & 192.18  & 213.16 & 45.08 &  55.81 & 0.06 & 33.80 
 \\
\cline{1-7} 
 Background & 154.76 & 88.06 & 30.95 &  36.58 & 6.40 & 40.32 
\\
\cline{1-7} 
\end{tabular}
\end{table*}

 \begin{figure}[]
    \centering
        \includegraphics[width=.4\linewidth]{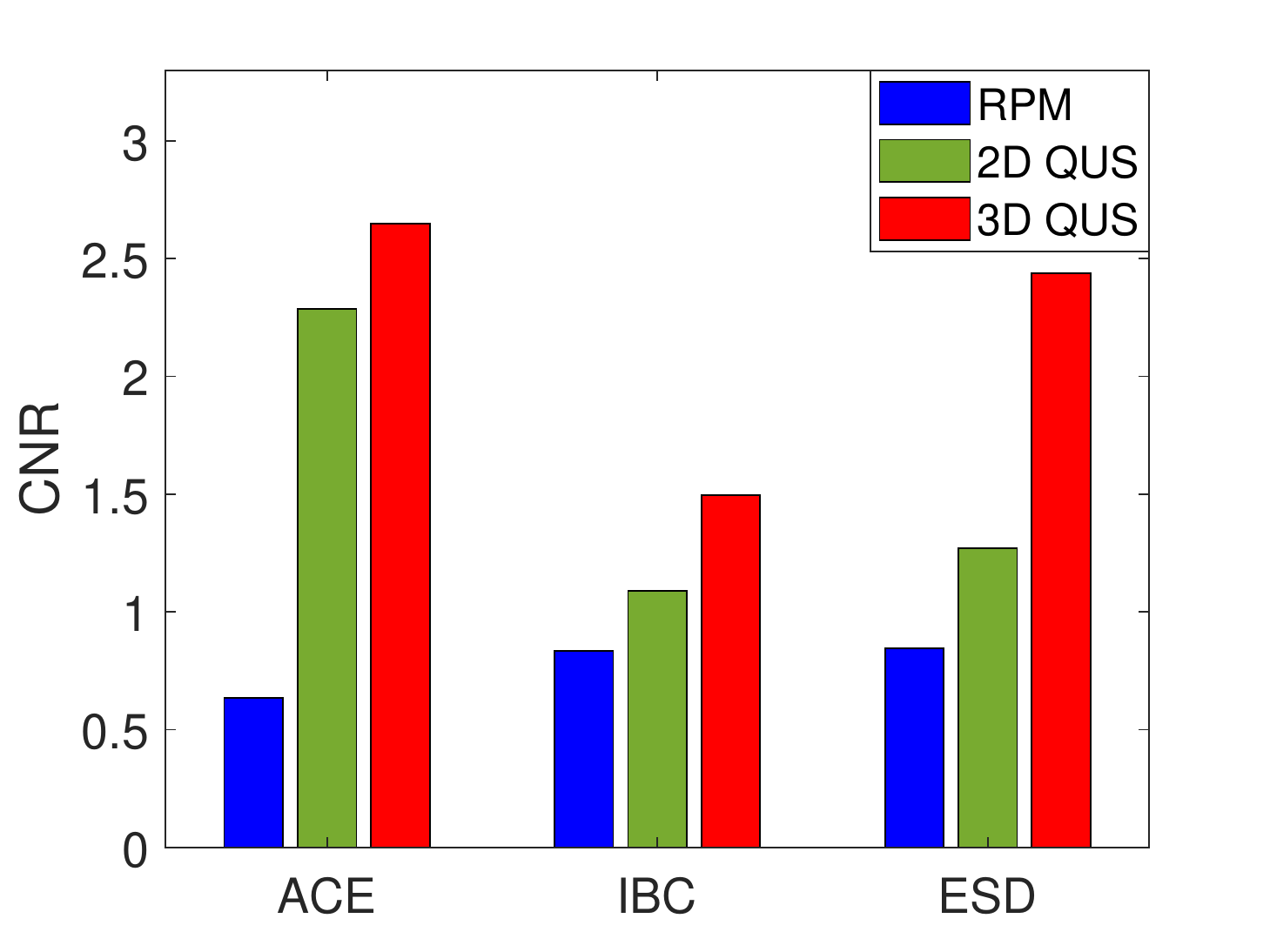}       
    \caption{CNR for ACE, IBC and ESD measures obtained from different methods.} 
    \label{figure:9}
    \end{figure}
    
\subsection{Effect of three-dimensional regularization} 
Unlike conventional QUS system, the proposed system in this work acquires multiple elevational planes of RF data. We perform a three-dimensional adaptive TV regularization to obtain the volumetric QUS maps. To study the effect of three dimensional regularization, we vary the number of planes and evaluate the precision by computing standard deviation of absolute percentage error for ACE,  IBC and ESD measures. We also investigate whether three dimensional regularization has any advantage over the two-dimensional counterpart. Therefore, we perform 2-dimensional adaptive regularization after averaging the estimated power spectra from multiple planes. The results have been shown in Fig. \ref{figure:10}. Also, the $1/\sqrt{N}$ curves for IBC and ESD have been superimposed on the graphs, where $N$ represents the number of planes used in the computation. For uncorrelated planes, the standard deviation of absolute percentage error curves should follow the $1/\sqrt{N}$ curve. The ACE variance remains unaffected (standard deviation of absolute percentage error< 0.5\%) as the number of planes increases for this particular example of homogeneous phantom. For IBC and ESD measures, increasing the number of planes improve the estimation variance for both 2D and 3D regularization methods. However, the 3D regularization not only outperforms the 2D regularization estimation precision, but the reduction in standard deviation of absolute percentage error measure improves over the $1/\sqrt{N}$ rate, which is the theoretical rate obtained by averaging $N$ uncorrelated realizations.  
 
 \begin{figure}[h]
    \centering
        \includegraphics[width=.4\linewidth]{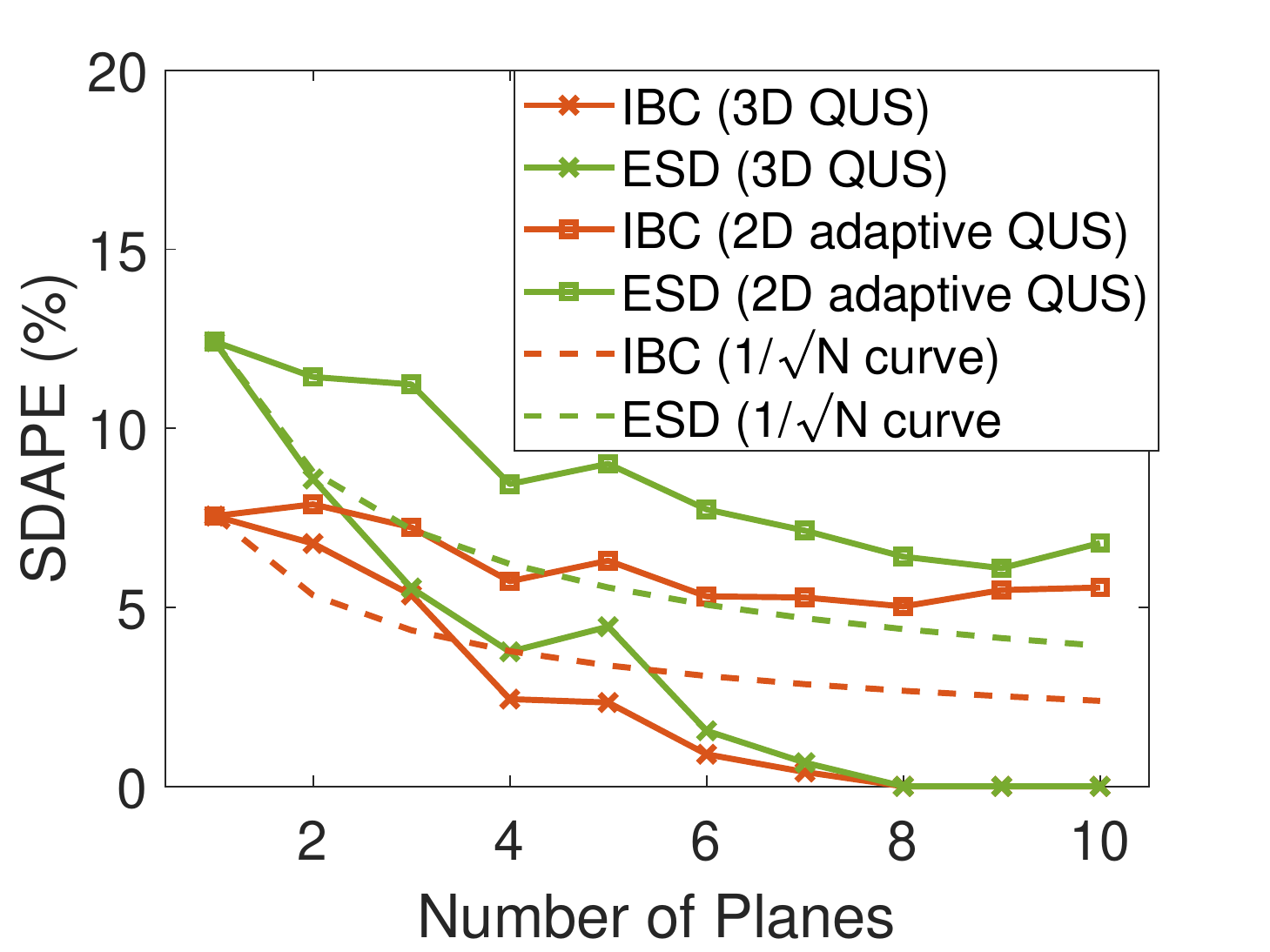}
    \caption{Standard deviation of absolute percentage error in IBC and ESD measure obtained using three-dimensional TV regularization and two-dimensional TV regularization vs. the number of planes used in the computation. The dashed lines represent the theoretical ($1/\sqrt N$) curves obtained by averaging $N$ completely uncorrelated planes. }
        \label{figure:10}
    \end{figure}
    
    \begin{figure}[h]
    \centering
        \includegraphics[width=.4\linewidth]{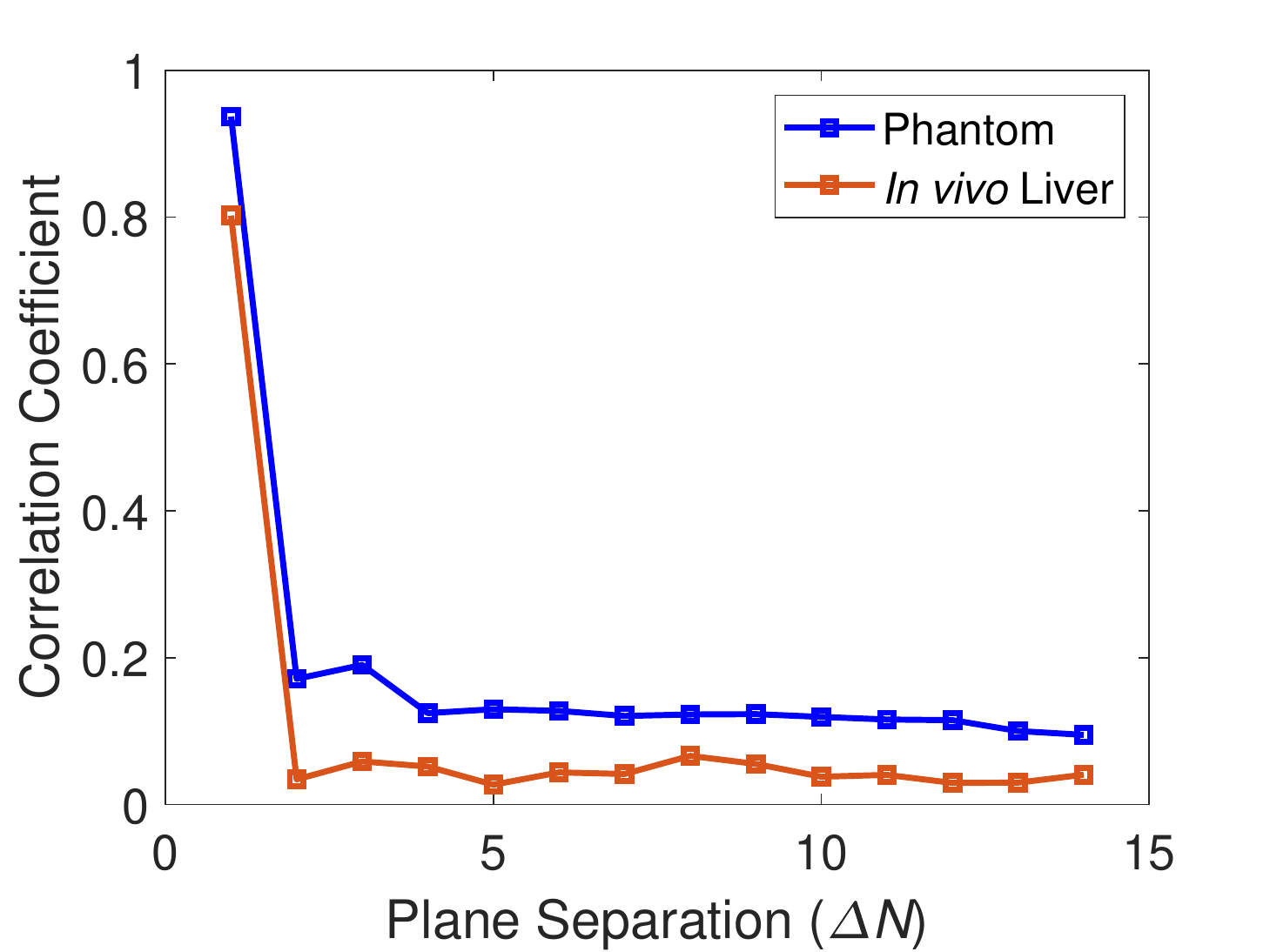}
    \caption{Correlation coefficient between two  2-D radio-frequency (RF) signals  separated by $\Delta N$ planes for a phantom and an \textit{in vivo} liver example. 
 }
        \label{figure:11}
    \end{figure}
    
    Figure \ref{figure:11} shows that the radio-frequency data obtained from \textit{in vivo} liver are more decorrelated compared to the one obtained from the phantom. The improved decorrelation between subsequent planes is an effective indicator of improvement in QUS measurement for \textit{in vivo} liver compared to phantom.

 \subsection{Effect of Spatial Weighting in Total Variation Regularization}  
 The phantom study demonstrates that the spatial weighting in the proposed 3D QUS method improves the QUS estimation, even in regions with high backscatter variation. In this section, we investigate the effect of spatial weighting on the selection of regularization parameter and QUS estimation performance. In \ref{figure:12}(a) and (b), we compare the cost function value for different values of regularization parameters obtained at the ground truth QUS measures for the 3D adaptive QUS and 2D uniform QUS method. As it is difficult to visualize a 4-dimensional cost function, we set the values of $\lambda_2$ and $\lambda_3$ equal. We found that the spatial weighting keeps the optimum cost comparatively less variable for different $\lambda_2$ and $\lambda_3$ values for variable ACE region-of-interest (homogeneous as dictated by low  $\Delta \mathit{SNR}^{env}$). The effect is more pronounced for phantom 3 with variable backscatter property (high $\Delta \mathit{SNR}^{env}$).
   
   Figure \ref{figure:12}(c) demonstrates that the error in QUS parameter estimation can be optimized by choosing a suitable regularization parameter set for both the 3D QUS and the 2D QUS method. However, for the 3D QUS, the optimum ranges for regularization parameter coincide, whereas for the 2D QUS with uniform regularization, the ranges do not overlap. Therefore, it is possible to select an optimum regularization parameter for the 3D QUS method to optimize QUS parameter estimation irrespective of homogeneity (backscatter variation). On the contrary, the 2D QUS with uniform regularization would require different parameters set depending on the underlying tissue characteristic, which would not be feasible for clinical application.    

   \begin{figure*}[]
    \centering
        \includegraphics[width=.8\linewidth]{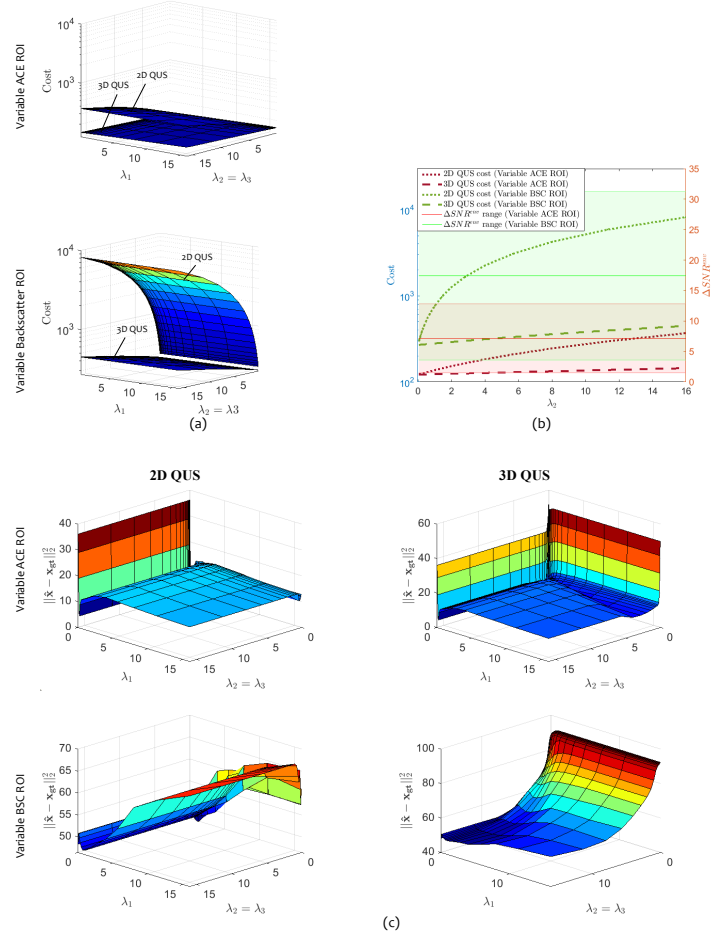}       
    \caption{Effect of spatial weighting using phantom 2 (variable ACE) and phantom 3 (variable backscatter and scatterer diameter property). (a) Cost evaluated at ground truth values of QUS paramters using the 2D QUS and the 3D QUS for different values of regularization parameters, $\lambda_1$, $\lambda_2$ and $\lambda_3$. For both cases, the upper surface is obtained from the 2D QUS and the lower one is from the 3D QUS. (b) Cost evaluated at ground truth values using the 2D QUS and the 3D QUS for different values of regularization parameter, $\lambda_2$, when $\lambda_1 = 2^{-4}$ and $\lambda_3 = \lambda_2$. The shaded regions indicate the range of $\Delta \mathit{SNR}^{env}$, where the mid line is the mean $\Delta \mathit{SNR}^{env}$ and the upper and lower lines are the standard deviation. (c) Error in QUS parameters estimation resulted from the 2D QUS and the 3D QUS for different values of regularization parameters, $\lambda_1$, $\lambda_2$ and $\lambda_3$  }
    \label{figure:12}
    \end{figure*}

\subsection{\textit{In vivo} Liver: Steatosis Detection Performance}    
\begin{figure}[h]
    \centering
        \includegraphics[trim={4.5cm 5cm 0 0cm},clip,scale=0.6]{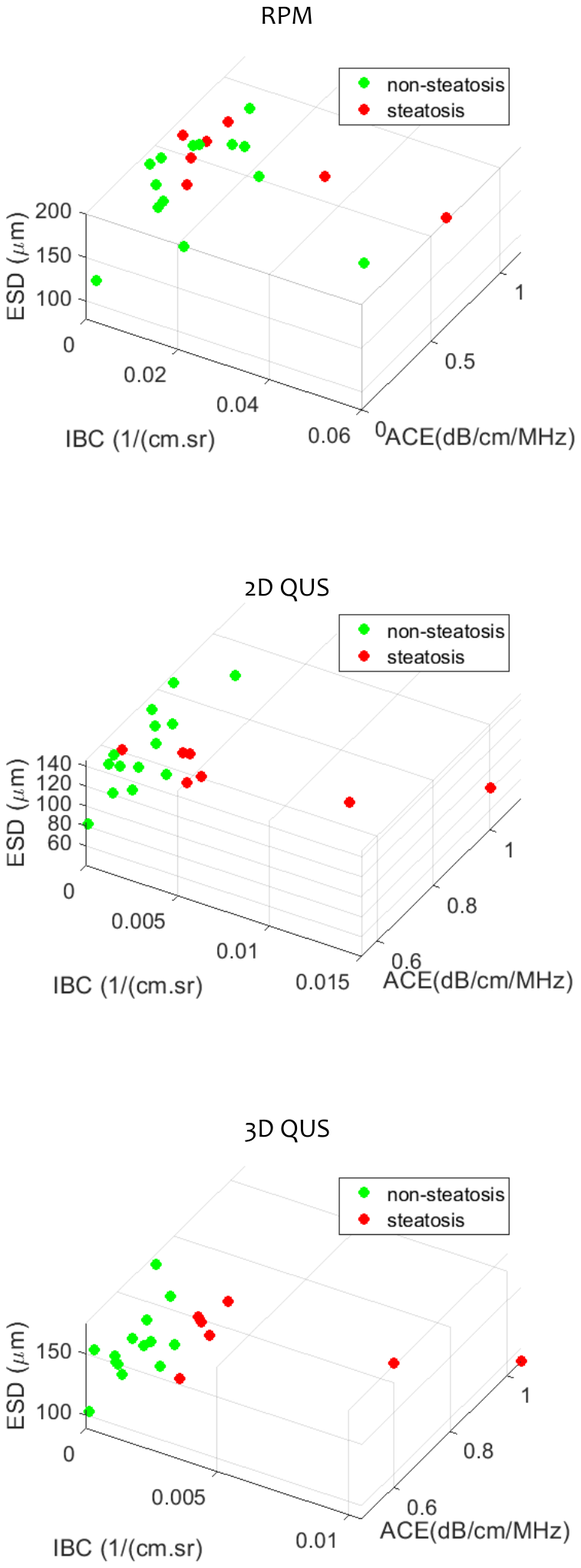}
    \caption{Scatter plot of QUS measures representing normal and steatosis cases.} 
    \label{figure:14}
    \end{figure}
\begin{figure*}[h]
    \centering
        \includegraphics[trim={0 11.2cm 0 1cm},clip,scale=0.7]{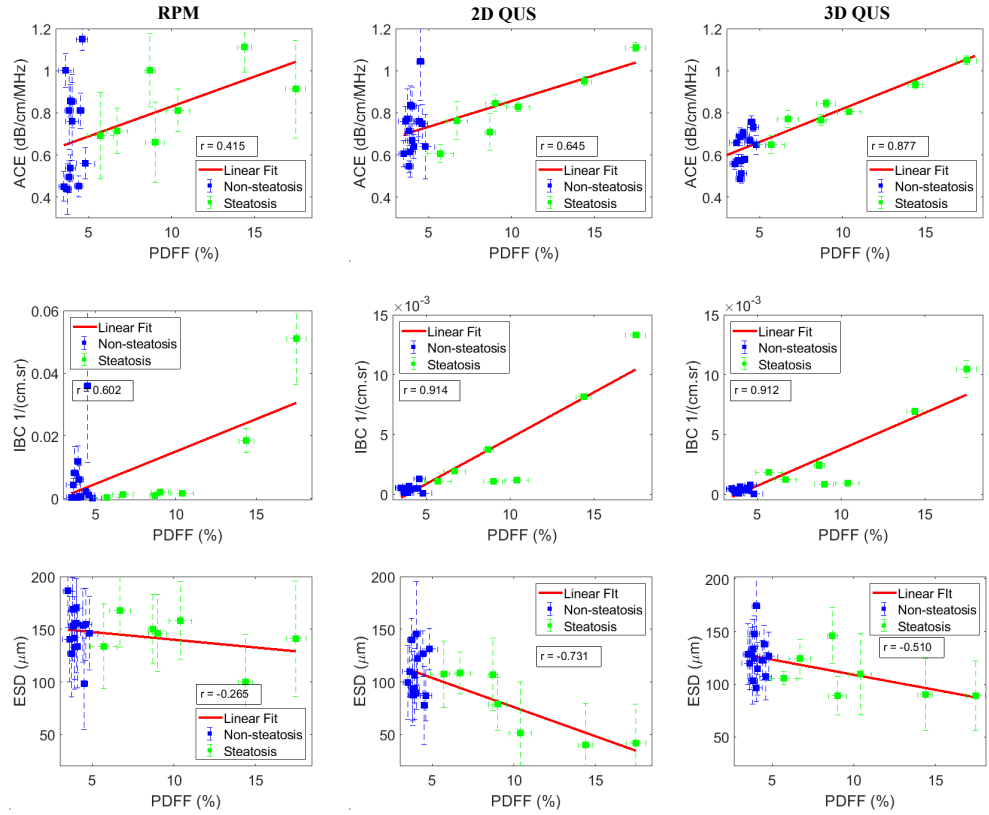}
    \caption{\textit{In vivo} human liver QUS measures from twenty-one patients and their correlation with proton density fat fraction. The vertical and horizontal error bars show the standard deviation in QUS measure and PDFF measure, respectively, whereas the square represents the mean calculated over a region-of-interest.} 
    \label{figure:13}
    \end{figure*}

    \begin{table*}[!t]
\caption{\label{tab4} Steatosis classification results for different QUS feature combination.}
\centering
\begin{tabular}{|c|c|c|c|c|c|c|}
\hline
\multirow{2}{*}{Features} & \multicolumn{3}{|c|} {Accuracy } & \multicolumn{3}{|c|} {AUROC} 
 \\
\cline{2-7} 
 & RPM & 2D QUS & 3D QUS &  RPM & 2D QUS & 3D QUS
\\
\cline{1-7} 
ACE &  0.62 &   0.71   & 0.95 & 0.70  &  0.68 &    0.93
\\
\cline{1-7} 
IBC &     0.67  &  0.81   & 0.95 &0.40&    0.97   & 1.00
 \\
\cline{1-7} 
ESD &0.67 &   0.81&    0.76&
0.53&    0.75&    0.76
\\
\cline{1-7} 
ACE \& IBC &     0.67  &  0.81 &  0.95&
    0.72&    0.97 &   1.00
\\
\cline{1-7} 
IBC \& ESD &      0.57   &  0.81   &  0.86& 
0.77  &   0.85 &    0.93
\\
\cline{1-7} 
ACE \& ESD &     0.67   &  0.81   &  0.95 &
0.52    &0.97    & 1.00
\\
\cline{1-7} 
ACE \& IBC \& ESD &     0.76  &  0.86   & 1.00&
   0.70    &0.97  & 1.00
\\
\cline{1-7} 
\end{tabular}
\end{table*}

The correlation results between MRI-PDFF and QUS measures have been displayed in \ref{figure:12}. The ACE and IBC measures demonstrated significant positive associations with MRI-PDFF, whereas ESD was negatively associated to MRI-PDFF. The proposed 3D QUS method yields stronger correlation between ACE and MRI-PDFF ($r=0.877, p<.001$) compared to the RPM ($r=0.415, p=0.062$) and the 2D QUS ($r=0.645, p= 0.002$) method. The IBC measure had similar and significantly strong association with MRI-PDFF for both the 2D QUS ($r=0.914, p<0.001$) and the 3D QUS method ($r=0.602, p<0.001$) and a moderate correlation for the RPM ($r=0.912, p= 0.004$). On the other hand, the 2D QUS ($r=-0.731, p<0.001$) and 3D QUS method ($r=-0.51, p=0.02$) yield strong to moderate negative association, whereas the noisy ESD results obtained by the RPM ($r=-0.265, p=0.53$) almost obscures the correlation.

  From the correlation analysis, the 2D QUS and the 3D QUS method exhibit comparable performance. However, a feature analysis plot including all three QUS features in Fig. \ref{figure:14} shows that the 3D QUS results in an improved separability between the steatosis and non-steatosis group compared to the 2D QUS method.
  
  We utilize a quadratic discriminant analysis (QDA) classification to classify the cases into steatosis and non-steatosis group. The classification accuracy and the area under the ROC curve (AUROC) were reported in Table \ref{tab4}. IBC, as a single feature, exhibits superior ability to discriminate while using the regularization methods: 2D QUS and 3D QUS. Among different feature combination, ACE and IBC, ACE and ESD, and all three feature combination yield comparable performance. 
    
 \section{Discussion}
 Adopting a total variation regularization scheme extends the trade-off between estimation accuracy, precision, and spatial resolution.  A window dimension of 30 scanlines (10 uncorrelated scanlines) and 35$\lambda$ (13 pulse lengths) was found to be the minimum reliable dimension without incorporating regularization. This findings lie within the range of 7-15 pulse lengths and 10-20 uncorrelated scanlines, which are presented as the dimensions offering the best trade-off between spatial resolution, accuracy, and precision \cite{rosado2013task} for QUS estimation. In contrast, incorporating the total variation regularization into the 2D and the 3D QUS methods results in a dramatic improvement in precision of both ACE and IBC, where standard deviation of absolute percentage error remains below $5\%$ for a window dimension of 10$\lambda$ ($\simeq 3.75$ pulse lengths) and 25$\lambda$ ($\simeq 9.40$ pulse lengths) axially and 5 ($\simeq2$ uncorrelated) scanlines, laterally. The mean of the absolute error of ACE and IBC remained below $6\%$ and $40\%$, respectively for both the regularization methods. We can compare the accuracy performance with the previous work \cite{vajihi2018} using a regularization method, which was able to achieve a mean percentage error (MPE) of 0.46\% in ACE and 327\% MPE  and 19.7\% MPE in estimation of backscatter parameters, $b$ and $n$. Our results remain consistent with the previous studies \cite{wear2005interlaboratory}, where backscatter estimations have been found to suffer from significant measurement error, ranging within 2 orders of magnitude. 
 
 	Both regularization methods yield similar performance for the homogeneous phantom (phantom 1) and variable ACE phantom (phantom 2), as can be seen in Fig. \ref{figure:5} and Fig. \ref{figure:6}. With a similar scatterer properties (including IBC and ESD), the $\Delta \mathit{SNR}^{env}$ values of both of these phantoms remain below $\Delta \mathit{SNR}^{env}_{min}$, with little effect of spatial weighting. Therefore, the uniform QUS and the weighted QUS methods are essentially similar when the tissue under experiment is homogeneous in terms of backscatter properties. The 3D QUS method outperforms the 2D QUS when the underlying tissue has variable backscatter properties, as is the case of phantom 3. Figure \ref{figure:6} shows that the $\Delta \mathit{SNR}^{env}$ map was able to identify the boundaries where backscatter properties vary. Incorporating the $\Delta \mathit{SNR}^{env}$ information in the 3D QUS enables adaptive regularization, and therefore results in more accurate reconstruction of the ACE and IBC map compared to the uniform regularization in 2D QUS. It should be noted that both the RPM and the 2D QUS method exhibit an underestimation and an overestimation of ACE centring the boundary between medium with different backscatter properties, a trend described in previous literature  \cite{pawlicki2013, Deeba2019a}. The 3D QUS method was able to overcome the issue, yielding accurate ACE and IBC results with close agreement to the values reported by the manufacturer.

For ESD computation, a Gaussian form factor has been used as it can successfully model the scattering properties of many soft tissues, including liver \cite{nguyen2019characterizing, Gilmore2008}. However, a spherical shell form factor is more appropriate for spherical glass beads, as were used in the phantoms in our experiments \cite{insana1990parametric, Gilmore2008}. The Gaussian form factor would consistently overestimate the sphere size, resulting in an effective diameter greater than the physical diameter \cite{insana1990parametric, Liu2010}. The effective diameter of the glass beads can be found by comparing the Gaussian form factor to the spherical shell form factor and computing the conversion factor for the particular physical diameter within the usable frequency range \cite{Liu2010}. The physical diameters for inclusion 1, inclusion 2, inclusion 3 and the background medium in phantom 3 are: 220 $\mathrm{\mu m} $, 120 $\mathrm{\mu m}$, 40 $\mathrm{\mu m}$, and 70 $\mathrm{\mu m}$, respectively. With conversion factors 1.27, 1.31, 1.56, and 1.28, respectively, the effective diameters have been found to be : 342, 158, 51, and 90 $\mu m$. The accuracy of ESD estimation depends on the estimation accuracy of the backscatter parameter, $n$. Therefore, the accurate backscatter parameter estimation obtained by the 3D QUS further is translated to accurate ESD estimation as can be seen in Fig. \ref{figure:7}.

The CIRS phantom was specifically designed to cover the wide range of QUS parameters to be expected in human liver \textit{in vivo} with different stage of steatosis. Figure \ref{figure:8}, it can be seen that the 2D QUS method and the 3D QUS method both reach close to the ground truth value with high precision for the entire range of ACE from 0.37 to 1.18 dB/cm/MHz, while both the mean and standard deviation of absolute percentage error remain below $10\%$. For IBC, the deviation of IBC from the ground truth values range from 0.90 to 3.63 dB using the 3D QUS method, where the mean absolute percentage error is below $25\%$ and standard deviation of absolute percentage error is below $11\%$.

The accuracy and precision of ESD estimation further depends on the range of $ka_{eff}$, the product of the wave number $k$ and the effective scatterer radius $a_{eff}$, which should be within 0.5 to 1.2 for optimum results \cite{Gilmore2008}. For the frequency range of 2 MHz to 3.5 MHz used in our experiments, the $ka_{eff}$ values for the scatterers in inclusion 1, 2, 3 and the background are: $ 1.40-2.44$, $0.64-1.13$, $0.21-0.36$, and $0.37-0.64$, respectively. Accordingly, we obtain high estimation error and variance for inclusion 1 and 3, as the corresponding $ka_{eff}$ values are well outside of the optimal range. The high variance in the ESD estimation for background scatterers can partly be attributed to the boundary artifacts. Finally, the $ka_{eff}$ for the scatterers in inclusion 2 is within the optimal range, resulting in an accurate and precise ESD estimation (Fig. \ref{figure:7}, Fig.\ref{figure:8}c and Table \ref{tab3}).

The advantage of using spatial weighting in the proposed 3D QUS over the uniform QUS and the baseline RPM is further evident from the CNR study. The 3D QUS improves the contrast significantly, with an average CNR improvement of $316\%$, $79\%$, and $188\%$ for ACE, IBC and ESD, respectively compared to the baseline RPM. Compared to the 2D QUS, the CNR improvement achieved using the 3D QUS method was $16\%$, $37\%$ and $92\%$ for ACE, IBC and ESD, respectively. This improved CNR would translate into better identification of abnormality in clinical application.

The proposed QUS imaging system utilizes the volumetric RF data acquired by a motorized 3D transducer. The use of a 3D system allows the volumetric QUS reconstruction, using a three-dimensional total variation regularization.Figure \ref{figure:10} shows that 3D regularization improves the estimation variance at a faster rate than the theoretical $1/\sqrt{N}$ curve obtained from averaging $N$ uncorrelated realizations. The advantage of utilizing volumetric data could be further enhanced in case of \textit{in vivo} liver data. The correlation between two radio-frequency signal data separated by $\Delta N$ planes, gives a measure of effectiveness of spatial compounding \cite{herd2011improving}, where a low normalized correlation coefficient value ($<0.2$) indicates independent data realizations. Figure \ref{figure:11} shows that the radio-frequency data obtained from \textit{in vivo} liver get decorrelated at a faster rate with the increasing plane separation compared to the one obtained from the phantom. As the ground truth values for QUS variance in \textit{in vivo} liver is unknown, we cannot obtain the exact measures of improvement with increasing data realizations. Nevertheless, the improved decorrelation between subsequent planes is an effective indicator of improvement in QUS measurement for \textit{in vivo} liver compared to phantom.

  The proposed 3D QUS method can be realized on a standard 3D ultrasound imaging system. This work was done as a part of a study together with S-WAVE. In a separate examination of data that repeated tests with and without external vibration for S-WAVE, the variability in QUS estimates were not found to be significantly different. Future work could further compare serial versus simultaneous S-WAVE and QUS acquisition.

Our QUS estimation results for liver \textit{in vivo} were comparable to the MRI-PDFF performance, the current gold standard. The presence of hepatic steatosis was defined as MRI-PDFF$>5\%$ \cite{caussy2018optimal}. From the correlation analysis, it was found that the ACE and IBC obtained using the proposed method have strong positive association with MRI-PDFF, whereas the ESD was found to be negatively correlated to MRI-PDFF (Fig. \ref{figure:13}). The ACE and IBC correlation results are in agreement with previous studies on human liver \textit{in vivo} \cite{andre2014accurate, tada2019utility}, additionally with an improved correlation. On the other hand, this is the first work to report the correlation between ESD of human liver and MRI-PDFF. Nevertheless, previous studies on rabbit fatty liver found negative correlation between ESD and fat content of the liver \cite{nguyen2019characterizing}. Using different combination of these three QUS parameters estimated using the proposed method with a simple QDA classifier, we were able to discriminate between the steatosis and the non-steatosis class with an accuracy up to $100\%$ and area under the receiver operating characteristic (AUROC) of 1. Among the QUS features, ESD was least effective to classify the two groups. However, the high contrast achievable using ESD (Fig. \ref{figure:9}) suggests that ESD, though exhibit inferior performance to identify steatosis which is a diffuse condition, might be effective in characterizing focal diseases.

Using Youden index, the optimal threshold for ACE, IBC and ESD to detect the presence of steatosis  were found to be 0.77 dB/cm/mHz, $8.5e^{-4}$ 1/(sr.cm), and 146.3 $\mu$m. The ACE and IBC cut-off agree with the previous findings, where the ACE and IBC cut-off values were equal to 0.76 dB/cm/MHz and $2.7e^{-3}$ 1/(sr.cm) \cite{andre2014accurate}. Among the limitations of our study is the ESD values obtained for steatosis participants, which partially lie within non-optimal $ka_{eff}<0.5$ range. Future work will involve application of comparatively higher frequency range to mitigate the problem. Also, we observe high variance in the ESD estimation for the liver study, which can be due to the presence of coherent component in the backscatter RF data. Therefore, we will further explore different signal decomposition techniques to separate the diffuse component from the coherent component \cite{nizam2020classification}. Also, instead of using a Gaussian form factor, tissue-specific form factors can be derived from the impedance maps of underlying tissue and could be used for improved ESD computation \cite{mamou2005identifying}. Another limitation of the \textit{in vivo} liver study is that the MRI-PDFF images and the ultrasound images are not registered. Though care has been taken to ensure that the planes-of-interest in MRI and ultrasound roughly coincide, a registration scheme would allow direct comparison with possible identification of focal lesions. 
 \section{Conclusion} 
 
 In this paper, we present a multiparametric 3D weighted QUS imaging system, which simultaneously measures three biomarkers for tissue characterization including ACE, IBC and ESD. The method incorporating a spatially adaptive regularization was able to achieve improved precision-resolution trade-off in QUS reconstruction in the presence of underlying tissue homogeneity. The three-dimensional regularization further improves the QUS estimation precision with a standard deviation reduction faster than the theoretical $1/\sqrt{N}$ rate achievable by averaging of $N$ uncorrelated realizations. With strong correlation to the current gold standard MRI-PDFF, each of the three parameters demonstrate the potential to detect hepatic steatosis. The combination of the QUS parameters further improve the steatosis classification performance. The proposed volumetric QUS system, built within an elastography system S-WAVE, therfore promises a comprehensive assessment tool for non-alcoholic fatty liver disease.

\section*{Acknowledgments}
The authors would like to thank the support from the Charles A. Laszlo Chair in Biomedical Engineering held by Professor Salcudean, Microsoft corporation, Schlumberger foundation, Canadian Institutes of Health Research (CIHR), and the Natural Sciences and Engineering Research 

\bibliography{ref}
\bibliographystyle{splncs04}

\end{document}